\documentclass[a4paper,onecolumn,superscriptaddress,nofootinbib,showpacs,amsmath,amssymb,aps,prd,notitlepage,floatfix]{revtex4-1}
\usepackage[T1]{fontenc}
\usepackage[latin1]{inputenc}
\usepackage{graphicx,amsfonts}
\usepackage{amsmath,xcolor}
\usepackage{amsthm}
\usepackage{amssymb}
\usepackage[hyperfootnotes=false]{hyperref}
\usepackage{cleveref}
\newcommand{\be}{\begin{equation}}
\newcommand{\ee}{\end{equation}}
\newcommand{\bea}{\begin{eqnarray}}
\newcommand{\eea}{\end{eqnarray}}
\newcommand{\bml}{\begin{subequations}}
\newcommand{\eml}{\end{subequations}}

\newcommand{\x}{\xi}

\newcommand{\ep}{\epsilon}

\newcommand{\bbm}{\begin{bmatrix}}
\newcommand{\ebm}{\end{bmatrix}}

\usepackage{physics}
\definecolor{burgundy}{rgb}{0.5, 0.0, 0.13}

\newcommand{\mcA}{{\mathcal{A}}}

\newcommand{\mcE}{{\mathcal{E}}}
\newcommand{\mcF}{{\mathcal{F}}}

\newcommand{\mcP}{{\mathcal{P}}}

\newcommand{\at}[2][]{#1|_{#2}}
\newcommand{\half}{\frac{1}{2}\,}
\newcommand{\inv}[1]{\frac{1}{#1}\,}

\newcommand{\christoffel}[3]{\Gamma^{#1}_{#2#3}}

\newcommand{\pd}[2]{\frac{\partial{#1}}{\partial{#2}}}
\def\eps{\varepsilon}
\newtheorem{theorem}{Theorem}[section]
\newtheorem{statement}[theorem]{Statement}

\begin{document}

\title{Causality and stability in first-order conformal anisotropic hydrodynamics}

\label{today}

\author{F\'abio S.\ Bemfica}
\affiliation{Escola de Ci\^encias e Tecnologia, Universidade Federal do Rio Grande do Norte, 59072-970, Natal, Rio Grande do Norte, Brazil}

\author{Mauricio Martinez}
\affiliation{Department of Physics, North Carolina State University, Raleigh, North Carolina 27695, USA}

\author{Masoud Shokri}
\affiliation{Institute for Theoretical Physics,
	Max-von-Laue-Strasse\ 1, D-60438 Frankfurt am Main, Germany}

\begin{abstract}
We formulate the first-order dissipative anisotropic hydrodynamical theory for a relativistic conformal uncharged fluid, which generalizes the Bemfica-Disconzi-Noronha-Kovtun first-order viscous fluid framework. Our approach maintains causal behavior in the nonlinear regime with or without general relativity coupling, and we derive and analyze the constraints on transport coefficients imposed by causality. We demonstrate the causal and stable behavior of our theory in specific cases, including the discussion of nonlinear causality as well as stability for linearized perturbations. We apply our newly developed first-order anisotropic theory to the Bjorken flow and show how causality and stability impose constraints on the behavior of the early-time attractor.
\end{abstract}

\maketitle
\section{Introduction}
Causality is one of the most important guiding principles in physics. In the bottom-up approach to constructing modern effective field theories, causality mandates that effective quantum field theories meet certain requirements, including microcausality, causal propagation limited to the speed of light, stability of the vacuum, and possible constraints on commutation relations \cite{deRham:2022hpx}. As other effective field theories, hydrodynamics is also subject to the restrictions imposed by causality. In this context, causality mandates that a solution to the relativistic fluid equations at a specific spacetime point $x$ is entirely defined by the past spacetime region that is causally connected to $x$ \cite{Hawking:1973uf,Wald:1984rg}. While causality is an important physical requirement for relativistic fluid dynamical equations of motion, it is not the only one. Two additional requirements are necessary for the equations of motion: to be locally well-posed and stable. The latter demands that small perturbations around the thermal state decay over time, while the former ensures that the system follows a well-defined spacetime evolution for a given set of initial conditions. The original first-order dissipative hydrodynamical equations of motion, known as the Navier-Stokes (NS) equations and derived  by Landau and Lifschitz \cite{landau2013fluid} as well as Eckart \cite{Eckart:1940te} showed to be acausal in the linear and nonlinear regimes \cite{Hiscock:1985zz,pichon}. To address the acausality and stability concerns of the Navier-Stokes equations, second-order hydrodynamics theories were introduced by Israel and Stewart (IS)\cite{Israel:1976tn,Israel:1979wp}. Since their seminal work, more recent formulations of second-order hydrodynamics have been developed \cite{Baier:2007ix,Denicol:2012cn,Hiscock:1983zz}. 

Within the realm of relativistic hydrodynamic theory and its phenomenological applications to different research areas, the absence of causality leads to the emergence of unphysical numerical solutions. The lack of causality in relativistic fluid dynamics prevents the equations from being well-posed, thereby making it impossible to guarantee the existence and uniqueness of solutions without causality. In a given numerical hydrodynamical simulation, while a specific reference frame may produce a numerical solution, its uniqueness cannot be guaranteed. Moreover, the solution may not even exist in a different reference frame. This highlights the critical and fundamental necessity to establish a causal theory for relativistic fluids.

A recent development in relativistic hydrodynamics is the BDNK theory proposed by Bemfica, Disconzi and Noronha \cite{Bemfica:2017wps,Bemfica:2019knx,Bemfica:2019hok,Bemfica:2020gcl,Bemfica:2020zjp} along with Kovtun \cite{Kovtun:2019hdm,Hoult:2020eho}. The BDNK theory has offered a practical and straightforward approach to address the long-standing issues of causality, stability, and local well-posedness in relativistic fluids.  Its development has created new opportunities for advancing our understanding of the fundamental principles underlying relativistic fluid dynamics. Basically BDNK theory is a generalization of the first-order NS relativistic theory which is causal in the linear and nonlinear regimes \footnote{The rigorous mathematical demonstrations of these statements are found in Refs. \cite{Bemfica:2019hok,Bemfica:2020gcl,Bemfica:2020zjp}.} as well as locally well-posed in Sobolev spaces in the presence and/or absence of gravity. BDNK theory is also linearly stable around global equilibrium. An interesting feature of BDNK theory is that the definition of the hydrodynamical variables, such as the energy density, receive out-of-equilibrium corrections. In this sense BDNK first-order theory differs from the standard approaches based on the Landau and Eckart frames.

It is essential to emphasize that the theoretical developments concerning causality extend far beyond merely seeking compatibility between causality and a particular physical theory. Indeed, recent independent studies conducted by two groups, Plumberg \textit{et al}.\ \cite{Plumberg:2021bme} and Chiu and Shen \cite{Chiu:2021muk}, have revealed violations of causality in various sized systems, including Pb-Pb and p+Au collisions, within state-of-the-art fluid dynamics simulations. These violations have distinct effects on different observables, pointing to the significant impact causality has on the outcomes.

On the other hand, anisotropic hydrodynamics has emerged as a very successful phenomenological model for describing nonequilibrium fluids with large spatial anisotropies  in high-energy nuclear collisions \cite{Martinez:2010sc,Florkowski:2010cf,Bazow:2013ifa,Molnar:2016gwq,Molnar:2016vvu} \footnote{For readers interested in delving into anisotropic hydrodynamics in heavy ions, we recommend the most recent review in Ref.\ \cite{Alqahtani:2017mhy}.}. It has been used to accurately model the spacetime evolution of the fireball in these collisions and has shown good agreement with experimental results (see Sec.\ 10 of Ref.\ \cite{Alqahtani:2017mhy} and references therein). Additionally, anisotropic hydrodynamics has passed rigorous numerical tests against exact kinetic theory models based on the Boltzmann equation (See Secs.\ 7-8 of Ref.\ \cite{Alqahtani:2017mhy} and references therein), providing confidence in its efficacy for studying fluids in far-from-equilibrium situations. Despite these successes, the foundational aspects of anisotropic hydrodynamics remain incompletely understood, particularly with respect to the role of causality and its constraints. Currently, anisotropic hydrodynamic simulations have displayed potential in handling realistic scenarios with considerable dissipative corrections. Nonetheless, the constraints imposed by causality, which might yield notable phenomenological consequences akin to those discussed in Refs. \cite{Plumberg:2021bme, Chiu:2021muk}, remain largely unexplored. Consequently, initiating a comprehensive investigation into the limitations imposed by causality in simulations of anisotropic hydrodynamics becomes a crucial and pressing task. On the other hand, there has been a growing interest in gaining a deeper understanding of the role of out-of-equilibrium corrections in simulations of neutron star mergers. It has been highlighted that these corrections can be significant \cite{Most:2021zvc, Alford:2020lla}, suggesting that standard hydrodynamical approaches may prove inefficient, as observed in heavy ion collisions. Given the demonstrated superior performance of anisotropic hydrodynamics in such scenarios, it becomes natural to explore its potential applications in the phenomenology of neutron star mergers. Hence, it becomes crucial to investigate how an anisotropic fluid behaves when coupled with gravity\footnote{Although some initial steps have been taken, it is important to note that these studies are still in their infancy \cite{Dash:2017uem}.}. Furthermore, advocates of anisotropic hydrodynamics have proposed its application in understanding protoquark and strange stars under the influence of strong magnetic fields \cite{Strickland:2012vu, Dexheimer:2012mk, Huang:2009ue}, where the direction of the magnetic field serves as the spatial anisotropy. Therefore, it is of utmost importance to analyze how causality constrains anisotropic hydrodynamics in scenarios distinct from heavy ion collisions. By doing so, we can gain valuable insights into the viability and limitations of this theory in diverse astrophysical environments.

Motivated by this gap in understanding, we introduce a novel approach in this work to investigate causality in anisotropic hydrodynamics. We develop a novel first-order theory for a conformal, uncharged fluid near thermal equilibrium that exhibits a spatial anisotropy characterized by a spacelike vector $l^\mu$. In addition, our work analyzes the role of causality when coupling anisotropic hydrodynamics with gravity, and thus, our findings pave the way for the inclusion of anisotropic dissipative effects in simulations of gravitational-wave signals coming from neutron star mergers.

The paper is organized as follows: in Sec.\ \ref{sec:abdnk} we discuss the most general form of a first-order anisotropic conformal fluid theory. We show in Sec.\ \ref{sec:entropy} that physical requirements of the second law of thermodynamics up to the order of validity of the theory leads unambiguously to the anisotropic BDNK theory of a conformal fluid. Section \ref{sec:causality} examines the conditions that causality imposes on the transport coefficients, both with and without gravity. Linear stability of our novel first-order anisotropic theory is analyzed in Sec.\ \ref{sec:stability}. The application of our novel anisotropic first-order theory for a fluid undergoing Bjorken flow is presented in Sec.\ \ref{sec:bjorken}. Our conclusions are summarized in Sec.\ \ref{sec:conclusions}. Technical details are found in the appendices. 

\paragraph*{Notations and conventions} We use the natural units in which $\hbar=c=k_B=1$, and adopt the Lorentzian metric $g_{\mu\nu}$ with signature $-+++$. The standard covariant derivative is denoted by $\nabla$, and the conformal covariant derivative by $\mathcal{D}$. We use the standard symmetrization and antisymmetrization notations, such that, for example, for a rank-2 tensor we obtain that $A_{\mu\nu}$ is $A_{(\mu\nu)} = \tfrac{1}{2}\left(A_{\mu\nu} +A_{\nu\mu}\right)$ and
$A_{[\mu\nu]} = \tfrac{1}{2}\left(A_{\mu\nu} -A_{\nu\mu}\right)$, respectively. The Riemann tensor is defined as 
$R^\sigma_{\rho\mu\nu} = 2\left(\partial_{[\mu}\christoffel{\sigma}{\nu]}{\rho}
+\christoffel{\sigma}{[\mu}{\beta}\christoffel{\beta}{\nu]}{\rho}\right)$, and the Ricci tensor as $R_{\mu\nu} = R_{\rho\mu\sigma\nu}g^{\rho\sigma}$.

\section{First-order anisotropic conformal BDNK theory}
\label{sec:abdnk}
A common approach in hydrodynamics is to expand physical observables in terms of gradients and truncate the series expansion at a given order in the derivatives. For example, if one considers the energy-momentum tensor $T^{\mu\nu}$, the gradient expansion takes the form $T^{\mu\nu}=\mathcal{O}(1)+\mathcal{O}(\partial)+\cdots$, where $\mathcal{O}(1)$ corresponds to the ideal fluid contributions while $\mathcal{O}(\partial^n)$ are the viscous corrections of order $n$ in derivatives (time and space derivatives written in covariant form) of the dynamical variables such as energy, and density. Therefore, first-order theories contain only the first-order derivative corrections $\mathcal{O}(\partial)$ \footnote{The gradient expansion is equivalent to the Knudsen expansion in kinetic theory \cite{Cercignani,degroot}.}. Since quantities such as temperature and number density are only unambiguously defined in equilibrium, their out-of-equilibrium corrections may differ due to different choices of $\mathcal{O}(\partial)$. In fact, different first-order theories are connected by transformations in these out-of-equilibrium variables \footnote{For example, consider a fluid's theory defined by a set of $N$ out-of-equilibrium thermodynamic variables ${\psi_a}$ with $a=1,..., N$, and suppose we perform a transformation $\psi_a\to \psi_a+\delta \psi_a$, where $\delta \psi_a$ is first order in the derivative of the thermodynamic variables $\psi_a$. This leads to the existence of two different first-order theories that are, in fact, equivalent up to the first-order corrections.}. The connection between the Landau-Lifshitz and Eckart frames, for instance, can be understood through such transformations \cite{Israel:1976tn}. Recent works have provided a comprehensive discussion on this subtle yet crucial aspect \cite{Kovtun:2019hdm,Bemfica:2020zjp,Dore:2021xqq}.

We shall develop the first-order theory of a conformal uncharged fluid which has a spatial anisotropy along an arbitrary direction determined by a spacelike vector $l^\mu$. We start by introducing the most general form of the energy-momentum tensor that describes this particular fluid.

 \subsection{Anisotropic energy momentum tensor}
 \label{subsec:emt}

Here, we turn to the problem of constructing a nonlinearly causal anisotropic extension of the conformal BDNK hydrodynamics for an uncharged fluid  \cite{Bemfica:2017wps}. The most general form for the energy-momentum tensor $T^{\mu \nu }$, decomposed in the directions parallel and orthogonal to the timelike velocity flow $u^\mu$ ($u^\mu u_\mu=-1$) and the anisotropic spacelike vector $l^\mu$ ($l^\mu l_\mu=1$) which is invariant under the little group $SO(2)$, reads as \cite{Molnar:2016gwq,Molnar:2016vvu}
\bea
\label{1}
T^{\mu \nu } &=&\mathcal{E}\, u^{\mu }u^{\nu }
+\mathcal{P}_l\, l^{\mu }l^{\nu }+\mathcal{P}_\perp\, \Xi ^{\mu \nu }+2\, M\, u^{\left( \mu \right. }l^{\left. \nu\right) }+2\, W_{\perp u}^{\left( \mu \right. }
u^{\left. \nu \right) }\nonumber\\
&&+2\, W_{\perp l}^{\left( \mu\right. }l^{\left. \nu \right) }
+\pi _{\perp }^{\mu \nu }\nonumber\\
&=&\mathcal{E}\, u^{\mu }u^{\nu }
+(\mathcal{P}_l-\mathcal{P}_\perp)\, l^{\mu }l^{\nu }+\mathcal{P}_\perp\, \Delta^{\mu \nu }+2\, M\, u^{\left( \mu \right. }l^{\left. \nu\right) }+2\, W_{\perp u}^{\left( \mu \right. }
u^{\left. \nu \right) }\nonumber\\
&&+2\, W_{\perp l}^{\left( \mu\right. }l^{\left. \nu \right) }
+\pi _{\perp }^{\mu \nu }\;.
\eea
Here, $\Delta^{\mu\nu}=g^{\mu\nu}+u^ \mu u^ \nu$ and $\Xi^{\mu\nu}=\Delta^{\mu\nu}-l^ \mu l^\nu$ are the projectors orthogonal to $u^\mu$, and to $u^\mu$ and $l^\mu$ together, respectively. In Eq.\ \eqref{1} we introduce the macroscopic quantities that correspond to different projections of the energy momentum tensor as follows
\begin{subequations}
\begin{eqnarray}
\mathcal{E}&=& u_\mu u_\nu T^{\mu\nu}\,,\\
M &= &T^{\mu\nu }u_{\mu }l_{\nu }\;,  \label{kinetic:M} \\
\mathcal{P}_{l} &=&T^{\mu \nu}l_{\mu }l_{\nu }\;,  \label{kinetic:P_l} \\
\mathcal{P}_{\perp } &=&\frac{1}{2}\, T^{\mu \nu }\Xi _{\mu \nu }\;, \label{kinetic:P_t}\\
W_{\perp u}^{\mu } &= &\Xi _{\alpha }^{\mu }T^{\alpha \beta }u_{\beta }\;, \label{kinetic:Wu_mu} \\
W_{\perp l}^{\mu } &=&-\Xi _{\alpha }^{\mu }T^{\alpha \beta }l_{\beta }\;,\label{kinetic:Wl_mu} \\
\pi _{\perp }^{\mu \nu } &= &
\frac{\left(\Xi^\mu_\alpha \Xi^\nu_\beta+\Xi^\mu_\beta \Xi^\nu_\alpha-\Xi^{\mu\nu}\Xi_{\alpha\beta}\right)}{2}\,T^{\alpha \beta }\equiv \Xi _{\alpha \beta }^{\mu \nu}T^{\alpha \beta }\;.  \label{kinetic:pit_munu}
\end{eqnarray}
\end{subequations}

Quantities $\mathcal{E}=\mathcal{P}_l+2\mathcal{P}_\perp,\,M,\,W_{\perp u}^\mu,\, W_{\perp l}^\mu$, and $\pi_\perp^{\mu\nu}$ contain four scalars, two vectors, and one tensor built out of the leading-order hydrodynamics fields and their first-order derivatives, with their concrete forms to be determined subsequently. In particular, $\mathcal{E}=\varepsilon+\mathcal{E}^{(1)}$, where $\mathcal{E}^{(1)}$ is the first-order correction to the energy and $\varepsilon$ is the equilibrium energy density. Assuming $T$ to be an effective temperature, then, in the conformal case, $\varepsilon=\varepsilon(T)\propto T^4$ and $\varepsilon=P_l+2P_\perp$, where  $P_l$ and $P_\perp$ are the leading-order contributions to $\mathcal{P}_l$ and $\mathcal{P}_\perp$, respectively. 
Due to the conformal invariance, the derivatives of the scalar fields are related through $\nabla_\mu P_{l}=(P_{l}/\varepsilon)\nabla_\mu \varepsilon$ and $\nabla_\mu P_{\perp}=(P_{\perp}/\varepsilon)\nabla_\mu \varepsilon$. 

In Eq.\ \eqref{1} the vectors $W_{\perp\,l}^\mu$, $W_{\perp\,u}^\mu$, and the symmetric traceless tensor $\pi^{\mu\nu}_\perp$ are orthogonal to $u$ and $l$, i.e.
\[
	V_\mu W_{\perp\,u}^\mu = V_\mu W_{\perp\,l}^\mu = V_\mu \pi^{\mu\nu}_\perp = 0\;,
\]
with $V=\{u,l\}$. Note that the above expressions lead to $\Xi_{\mu\nu}\pi_\perp^{\mu\nu}=0$.  In the absence of other conserved currents, the fluid's evolution is governed by the energy-momentum conservation $\nabla_\nu T^{\mu\nu}=0$, which will be referred to as the equations of motion (EOM) and due to the conformal invariance can be equivalently expressed as $\mathcal{D}_\nu T^{\mu\nu}=0$, where $\mathcal{D}_\mu$ is the conformal covariant derivative compatible with $u^\mu$ \cite{Loganayagam:2008is}. In particular,
\begin{equation*}
	\mathcal{D}_\mu u_\nu=\Delta^\alpha_\mu\nabla_\alpha u_\nu-\Delta_{\mu\nu}\nabla_\alpha u^\alpha/3\,
\end{equation*}
and
\begin{equation*}
	\mathcal{D}_\mu\varepsilon=\nabla_\mu \varepsilon+4\varepsilon(u^\alpha\nabla_\alpha u_\mu-u_\mu \nabla_\alpha u^\alpha/3)\;.
\end{equation*}
The EOM may be decomposed into the directions parallel to $u$ and $l$ and the directions perpendicular to both as
\bml
\label{EOM}
\bea
u^\alpha \mathcal{D}_\alpha \mathcal{E}&=&-(\mathcal{P}_l-\mathcal{P}_\perp)l^\mu l^\nu \sigma_{\mu\nu}-M \mathcal{D}_\alpha l^{\alpha}
-l^\nu\mathcal{D}_\nu M-\mathcal{D}_\nu W^\nu_{\perp u}-2W_{\perp l}^\nu l^{\alpha}\sigma_{\alpha\nu}\nonumber\\
&&-\pi _{\perp }^{\mu\nu}\sigma_{\perp \mu\nu},\label{EOMa}\\
l^\alpha\mathcal{D}_\alpha \mathcal{P}_l&=&-(\mathcal{P}_l-\mathcal{P}_\perp) \mathcal{D}_\alpha l^{\alpha }- M\,l^{\mu}l^\nu\sigma_{\mu\nu}-u^\alpha\mathcal{D}_\alpha M-l_\nu W_{\perp u}^{\alpha}\mathcal{D}_\alpha u^{\nu}-\,l_\nu u^{\alpha}\mathcal{D}_\alpha W_{\perp u}^{\nu}\nonumber\\
&&-l_\mu l^\nu\mathcal{D}_\nu W^\mu _{\perp l}-\mathcal{D}_\alpha W_{\perp l}^{\alpha}
+\pi _{\perp }^{\nu \alpha } \mathcal{D}_\alpha l_\nu,\\
\Xi^{\mu\alpha}\mathcal{D}_\alpha \mathcal{P_\perp}&=&-(\mathcal{P}_l-\mathcal{P}_\perp)\,\Xi^\mu_\nu l^\alpha \mathcal{D}_\alpha l^\nu- M\,\Xi^\mu_\nu l^{\alpha}\mathcal{D}_\alpha u^{\nu}-\,M\Xi^\mu_\nu u^{\alpha}\mathcal{D}_\alpha l^{\nu} -\Xi^\mu_\nu W_{\perp u}^{\alpha}\mathcal{D}_\alpha u^{\nu}\nonumber\\
&&-\Xi^\mu_\nu u^{\alpha}\mathcal{D}_\alpha W_{\perp u}^{\nu}
- W_{\perp l}^{\mu}\mathcal{D}_\alpha l^{\alpha}-\Xi^\mu_\nu W_{\perp l}^{\alpha}\mathcal{D}_\alpha l^{\nu}-\Xi^\mu_\nu l^{\alpha}\mathcal{D}_\alpha W_{\perp l}^{\nu}
-\Xi^\mu_\nu\mathcal{D}_\alpha\pi _{\perp }^{\nu \alpha },
\eea
\eml
where $\sigma_{\mu\nu}=\mathcal{D}_{(\mu}u_{\nu)}=\Delta^{\alpha\beta}_{\mu\nu}\nabla_\mu u_\nu$ is the shear tensor, with $\Delta^{\alpha\beta}_{\mu\nu}=[\Delta^\alpha_\mu\Delta^\beta_\nu+\Delta^\beta_\mu\Delta^\alpha_\nu-(2/3)\Delta^{\alpha\beta}\Delta_{\mu\nu}]/2$. We have also introduced a transverse shear tensor as
\begin{equation*}
	\sigma_{\perp \mu\nu}=\Xi^{\alpha\beta}_{\mu\nu}\mathcal{D}_\alpha u_\beta=\Xi^{\alpha\beta}_{\mu\nu}\nabla_\alpha u_\beta\;.
\end{equation*}

It is important to keep in mind that an expression that only includes derivatives up to a certain order may be approximated at that order when the equations of motion are considered. As an example, one can rewrite \eqref{EOMa} by separating the leading- and first-order contributions from $\mathcal{E}$ and $\mathcal{P}$ to obtain
\bea
\label{EOMa2}
u^\alpha \mathcal{D}_\alpha \varepsilon+(P_l-P_\perp)l^\mu l^\nu \sigma_{\mu\nu}&=&-u^\alpha \mathcal{D}_\alpha \mathcal{E}^{(1)}-(\mathcal{P}_l^{(1)}-\mathcal{P}_\perp^{(1)})l^\mu l^\nu \sigma_{\mu\nu}-M \mathcal{D}_\alpha l^{\alpha}-l^\alpha\mathcal{D}_\alpha M-\mathcal{D}_\nu W^\nu_{\perp u}\nonumber\\
&&-\,2 W_{\perp l}^{\alpha}l^\nu\sigma_{\mu\nu}-\pi _{\perp }^{\mu\nu}\sigma_{\perp \mu\nu}.
\eea
Upon comparing the left- and right-hand sides of the equation above, it becomes evident that the terms $u^\alpha \mathcal{D}_\alpha \varepsilon$ and $l^\alpha l_\nu \mathcal{D}_\alpha l^\nu$ only contain first-order derivatives of the zeroth-order quantities. However, the combination $u^\alpha \mathcal{D}_\alpha \varepsilon+(P_l-P_\perp)l^\mu l^\nu \sigma_{\mu\nu}$ is second-order \textit{on-shell}, meaning that it becomes second order when the equations of motion given by \eqref{EOMa2} are taken into account. It can be said that the same combination is of the first-order \textit{off-shell}. Similar arguments may be applied in the remaining equations of motion in \eqref{EOM} above. In the next section, we make use of generic arguments based on the second law of thermodynamics together with the above \textit{on-shell} analysis in order to determine the remaining first-order terms $M$, $E^{(1)}$, etc., by assuming the physical applicability of a first-order theory.

\section{Entropy and entropy production}
\label{sec:entropy}

The second law of thermodynamics mandates that entropy should be at a maximum in a state of equilibrium, and the entropy production should not have a negative value. However, for a first-order theory one must only consider the entropy up to the first-order on-shell derivative, and hence the entropy production up to the second-order on-shell \cite{Kovtun:2019hdm}. This ensures that the theory is being applied outside of its intended physical applicability.

The entropy current of an anisotropic conformal uncharged fluid reads as (see Appendix \ref{Appendix_A})
\be
\label{Entropy_current}
S^\mu=\frac{P_\perp u^\mu-u_\nu T^{\mu\nu}}{T}\;,
\ee
where $P_\perp$ is the leading-order of $\mathcal{P}_\perp$, i.e., $\mathcal{P}_\perp=P_\perp+\mathcal{P}_\perp^{(1)}$. In this case, the entropy density reads
\be
\label{Entropy}
-u_\mu S^\mu=s+\frac{\mathcal{E}^{(1)}}{T}\;,
\ee 
where the leading order contribution to the entropy density is $s=(\varepsilon+P_\perp)/T$. From Eq.\ \eqref{Entropy_current} we calculate the entropy production 
\bea
\label{Entropy_production}
\nabla_\mu S^\mu &=&-\frac{\pi_\perp^{\mu\nu} \sigma_{\perp \mu\nu}}{T}-\frac{2W_{\perp l}^{\mu}l^{\nu}\sigma_{\mu\nu}}{T}-\frac{\varepsilon-3P_\perp}{4T}\frac{u^\alpha\mathcal{D}_\alpha\varepsilon}{\varepsilon}-\frac{(\mathcal{P}_l-\mathcal{P}_\perp)l^\mu l^\nu \sigma_{\mu\nu}}{T}\nonumber\\
&&-\frac{\mathcal{E}^{(1)}u^\nu+W_{\perp u}^\nu+M\,l^\nu}{4T}\frac{ \mathcal{D}_\nu \varepsilon}{\varepsilon}\;.
\eea
The choices $\pi_{\perp\,\mu\nu}\propto -\sigma_{\perp\,\mu\nu}$ and $W_{\perp\,l}^\mu\propto -\Xi^{\mu\nu}l^\alpha \sigma_{\alpha\nu}$ give positive contributions to the first two terms in \eqref{Entropy_production}. On the other hand, the third term
\begin{equation}
    -\frac{\varepsilon-3P_\perp}{4T}\frac{u^\alpha\mathcal{D}_\alpha\varepsilon}{\varepsilon}
    \label{eq:thirdterm}
\end{equation}
cannot be positive definite since $\mathcal{D}_\alpha\varepsilon$ has no definite sign. Furthermore, the order of derivatives of this term, as given by Eq.\ \eqref{EOMa2}, depends on the leading order of pressures and could be either first or second order. 
Equation \eqref{eq:thirdterm} becomes second-order on-shell only if $P_\perp=P_l$. 
Hence, to ensure a non-negative entropy production, this term must be eliminated by letting $P_\perp=\varepsilon/3$.  
On the other hand, the term 
\begin{equation}
    \label{eq:anotherterm}
	\frac{(\mathcal{P}_l-\mathcal{P}_\perp)l^\mu l^\nu \sigma_{\mu\nu}}{T}
\end{equation}
is of first order in derivatives by assuming that there are only on-shell contributions in Eq.\ \eqref{EOM} and $\sigma_{\mu\nu}$ is of the first order in derivatives. 
Accordingly, this term has no definite sign and must be eliminated as well. 
To this end, the leading orders of the pressures must be equal
\be\label{eq:euil}
P_\perp=P_l=P=\frac{\varepsilon}{3}\;,
\ee
which we consider to be the case hereafter. Subsequently, and by following the prescription outlined in the previous section, the simplest anisotropic extension of the conformal BDNK hydrodynamics of an uncharged fluid is formulated by the energy-momentum tensor
\bea
\label{EMT}
T^{\mu\nu}&=&(\varepsilon+\mathcal{E}^{(1)})\, u^{\mu }u^{\nu }
+(\mathcal{P}_l^{(1)}-\mathcal{P}_\perp^{(1)})\, l^{\mu }l^{\nu }+(P+\mathcal{P}_\perp^{(1)})\, \Delta^{\mu \nu }+2\, M\, u^{\left( \mu \right. }l^{\left. \nu\right) }+2\, W_{\perp u}^{\left( \mu \right. }
u^{\left. \nu \right) }\nonumber\\
&&+2\, W_{\perp l}^{\left( \mu\right. }l^{\left. \nu \right) }
+\pi _{\perp }^{\mu \nu }\;,
\eea
where its components are written as
\bml
\label{Definitions}
\bea
\mathcal{E}^{(1)}&=&\frac{3\chi}{4}\frac{u^\mu\mathcal{D}_\mu\varepsilon}{\varepsilon}\;,\\
\mathcal{P}_l^{(1)}&=&\frac{\chi_l}{4}\frac{u^\mu\mathcal{D}_\mu\varepsilon}{\varepsilon}-2\eta_{ll}l^\alpha l^\beta\sigma_{\alpha\beta}\;,\label{def_b}\\
\mathcal{P}_\perp^{(1)}&=&\frac{\chi_\perp}{4}\frac{u^\mu\mathcal{D}_\mu\varepsilon}{\varepsilon}+\eta_{ll}l^\alpha l^\beta\sigma_{\alpha\beta},\label{def_c}\\
\pi_\perp^{\mu\nu}&=&-2\eta_\perp \sigma_\perp^{\mu\nu}\;,\\
W_{\perp l}^\mu&=&-2\eta_l \Xi^\mu_\lambda l_\nu \sigma^{\lambda\nu}\;,\\
W_{\perp u}^\mu&=&\frac{\lambda_\perp }{4}\Xi^{\mu\nu}\frac{\mathcal{D}_\nu \varepsilon}{\varepsilon}\;,\\
M&=&\frac{\lambda_l}{4}\frac{l^\nu \mathcal{D}_\nu \varepsilon}{\varepsilon}\;.
\eea
\eml
In the previous expression we have eight transport coefficients $
\{\chi,\chi_l,\chi_\perp,\eta_{l},\eta_{ll},\eta_{\perp},\lambda_l,\lambda_\perp\} 
$ which are proportional to $\varepsilon^{3/4}$, due to the conformal invariance. The conformal invariance also requires $\chi_l+2\chi_\perp=3\chi$ to ensure $T^\mu_\mu=0$.

Plugging Eqs. \eqref{Definitions} into the EOMs \eqref{EOM} and assuming condition \eqref{eq:euil}, the entropy density is given by $-u_\mu S^\mu=(\varepsilon+P)/T+\mathcal{O}(\partial^2)$  and the on-shell entropy production is simply 
\be
\nabla_\mu S^\mu=\frac{\pi_\perp^{\mu\nu} \pi_{\perp \mu\nu}}{2\eta_\perp T}+\frac{W_{\perp l}^\mu W_{\perp l\, \mu}}{\eta_l T}
+\frac{\left (\mathcal{P}_\perp^{(1)}-\mathcal{P}_l^{(1)}\right )^2}{3\eta_{ll}T}+\mathcal{O}(\partial^3)\;.
\ee  
We conclude that the on-shell entropy production given by the previous expression is positive up to second order in derivatives if 
\begin{equation}\label{eq:onshell-positivity}
	\eta_\perp \geq 0 \;, \qquad \eta_{ll} \geq 0\;, \qquad \eta_{l} \geq 0\;.
\end{equation}

\subsection{The isotropic conformal limit of BDNK}
\label{subsec:isoBDNK}
Once the spacelike anisotropic vector $l$ is chosen, the isotropic limit for the conformal uncharged fluid  is reproduced by setting
\begin{equation} \label{L1}
	\eta_\perp=\eta_l=\eta_{ll}=\eta\;,\qquad
	\chi_l=\chi_\perp=\chi\;,\qquad
	\lambda_\perp=\lambda_l=\lambda\;.
\end{equation} 
If we plug the above condition into \eqref{EMT}, the energy-momentum tensor reduces to the one derived in BDNK theory \citep{Bemfica:2017wps}, i.e., 
\bea
T^{\mu\nu}&=&\left [\varepsilon+\frac{3\chi}{4}\frac{u^\mu\mathcal{D}_\mu\varepsilon}{\varepsilon} \right ]u^\mu u^\nu+\left [P+\frac{\chi}{4}\frac{u^\mu\mathcal{D}_\mu\varepsilon}{\varepsilon}\right ]\Delta^{\mu\nu}-2\eta\sigma^{\mu\nu}+\frac{\lambda u^{(\mu}\Delta^{\nu)\alpha}}{2}\frac{\mathcal{D}_\alpha \varepsilon}{\varepsilon}\;,
\label{eq:bdnklim}
\eea
where $\chi,\lambda,\eta\propto \varepsilon^{3/4}$. The equivalence between the expressions derived above  and the corresponding one in BDNK theory \cite{Bemfica:2017wps} becomes clearer when considering the limit established by Eqs.\eqref{L1} while writing $\Delta^{\mu\nu}=\Xi^{\mu\nu}+l^\mu l^\nu$ in Eq.\ \eqref{eq:bdnklim}
together with the following identity \cite{Molnar:2016gwq,Molnar:2016vvu}
\bea
\sigma_{\mu\nu}&=&\sigma_{\perp \mu\nu}-\frac{1}{2}\Xi_{\mu\nu}l^\alpha l^\beta\sigma_{\alpha\beta}+2\Xi^{\alpha}_{(\mu}l_{\nu)}l^\beta \sigma_{\alpha\beta}+l_\mu l_\nu l^\alpha l^\beta \sigma_{\alpha\beta}.
\eea
Note that the traceless components of the energy momentum tensor \eqref{eq:bdnklim} are $\sigma_{\perp \mu\nu}$, $\Xi^{\alpha}_{(\mu}l_{\nu)}l^\beta \sigma_{\alpha\beta}$, and the combination $\Xi_{\mu\nu}l^\alpha l^\beta\sigma_{\alpha\beta}-2l_\mu l_\nu l^\alpha l^\beta \sigma_{\alpha\beta}$. 
These terms are needed and explain why these multiply the same coefficient $\eta_{ll}$ in Eqs.\ \eqref{def_b} and \eqref{def_c}.

\section{Causality}
\label{sec:causality}

Now, we shall show that there exist conditions for the transport parameters defined in Eqs. \eqref{Definitions} such that the resulting hydrodynamic theory exhibits nonlinear causality. It is worth noting that the EOM $\nabla_\nu T^{\mu\nu}=0$ give rise to a system of quasilinear partial differential equations (PDE), i.e., PDEs that do not contain products of their highest-order derivative terms. The independent variables in the energy-momentum tensor include the energy density $\eps$ and the components of the fluid's velocity $u^\mu$, with the anisotropic vector $l^\mu$ \footnote{The components of the vector $l^\mu$ are constrained by being a unitary spacelike vector $l^\mu l_\mu=1$ and being orthogonal to $u^\mu$. For instance, at the earliest stages of a heavy-ion collision, it is common to choose $u^\mu=\gamma(v_z)(1,0,0,v_z)$ with $l^\mu=\gamma(v_z)(v_z,0,0,1)$ \cite{Molnar:2016vvu}.} being chosen accordingly. To investigate causality in quasilinear systems, we analyze the principal part of the system of equations. This part contains only the terms of the highest order in each variable and determines the order of the partial differential equations in each variable. As an example, a quasilinear system that is of order one in $\eps$ and two in $u^\mu$ does not contain terms such as $(\partial\varepsilon)^2$, $(\partial^2 u)^ 2$, or $\partial\varepsilon\,\partial^2 u$, and its principal part can contain only terms of form $\partial \varepsilon$ or $\partial^2 u$. As explained in \cite{Bemfica:2017wps}, we assume $u^\mu$ to have four independent components, with the constraint $u^\mu u_\mu = -1 $ being imposed at an initial time and being preserved through the fluid's evolution. The EOM is then decomposed in directions parallel and perpendicular to $u^\mu$. This gives rise to five equations of the five independent variables 
\be
\label{Projected_EOM}
-u_\mu \nabla_\nu T^{\mu\nu}=0,\quad \Delta^\mu_\alpha\nabla_\nu T^{\alpha\nu}=0\;,
\ee 
with the constraint $u^\mu u_\mu = -1$ being used when required \footnote{This is equivalent of considering $\nabla_\nu T^{\mu\nu}=0$ together with the evolution equation for the constraint $u^\beta\nabla_\beta [u^\alpha\nabla_\alpha(u^\mu u_\mu)]=0$.}. Let us now consider Eqs. \eqref{Projected_EOM} together with the constitutive relation \eqref{EMT} for the energy-momentum tensor, and the dissipative fluxes given in \eqref{Definitions}, coupled with gravity through Einstein's equation 
\be
\label{GR_eq}
R^{\mu\nu}-\frac{1}{2}g^{\mu\nu}R=8\pi G T^{\mu\nu}\;,
\ee
and the gauge freedom fixed by assuming the harmonic gauge $g^{\mu\nu}\Gamma_{\mu\nu}^\alpha=0$.  As explained in Appendix \ref{app:causality}, the causality of the system is determined by the vectors $\xi_\mu = \nabla_\mu \Phi$ normal to the characteristic hypersurface $\Phi(x)=0$, which are the roots of the characteristic equation \eqref{h}. With the roots of the characteristic equation obtained as $\xi_0 = \xi_0(\xi_i)$, the system is causal if  (C1) they are real,
\begin{equation}\label{C1}
	\xi_0 \in \mathbb{R}\;,
\end{equation}
and (C2) $\xi_\alpha = (\x_0,\xi_i)$ is not timelike, i.e., 
\begin{equation}\label{C2}
	\xi_\alpha \xi^\alpha \geq 0\;.
\end{equation}
In our case, the characteristic equation has 30 roots, of which 20 are lightlike and thus causal, arising from pure gravity. The remaining pieces of the characteristic equation are spacelike roots, which we refer to as the matter sector, and can be further decomposed into two parts. The first part, which contains 4 roots, is
\be
\label{matter1}
A^2=\left (\lambda_\perp a^2-\eta_\perp v^2+\delta\eta_{\perp l} b^2\right )^2=0\;,
\ee
where 
\begin{equation}
	\delta\eta_{\perp l} = \eta_{\perp} - \eta_{l}\;,
	\qquad 
	a = u^\mu \xi_\mu\;,
	\qquad
	b = l^\mu \xi_\mu\;,
	\qquad
	v^\mu = \Delta^{\mu\nu}\xi_\mu\;.
\end{equation}
The second part which contains the 6 remaining roots reads as
\bea
\label{matter2}
H^\parallel_\parallel(V,\xi)\left [A^2+A\left (U_1^\mu l_{\mu}+U_2^\mu \xi_{\mu}\right )+U_1^\mu l_{\mu}U_2^\nu \xi_{\nu}-U_1^\mu \xi_{\mu}U_2^\nu l_{\nu}\right ]=0\;.
\eea
The explicit forms of the terms $H^\parallel_\parallel(V,\xi)$, $U_1^\mu$, and $U_2^\mu$ are calculated explicitly in Appendix \ref{app:causality}, which for convenience we listed below 
\bml
\bea
&&U_1^\mu l_\mu =b^2\delta\eta_{lll\perp} +a^2\delta\lambda+\delta\eta_{\perp l} v^2 +b^2\delta\eta_{lll}-\frac{a^2b^2\delta\lambda(\chi_\perp+\lambda_\perp+\delta\lambda+\delta\chi) }{4\varepsilon H^\parallel_\parallel(V,\xi)}\;,\\
&&U_1^\mu\xi_\mu =b^3\delta\eta_{lll\perp} +a^2b\delta\lambda+(\delta\eta_{\perp l}+\delta\eta_{lll})b v^2-\frac{a^2b\delta\lambda[(\chi_\perp+\lambda_\perp)v^2 +b^2(\delta\lambda+\delta\chi)]}{4\varepsilon H^\parallel_\parallel(V,\xi)}\;,\\
&&U_2^\mu l_\mu =b\left (\frac{\delta\chi}{3}+\delta \eta_{lll}\right )+\frac{(\chi_\perp-\eta_{ll})b}{3}-\frac{a^2b (\chi+\lambda_\perp) (\chi_\perp+\lambda_\perp+\delta\lambda+\delta\chi) }{4\varepsilon H^\parallel_\parallel(V,\xi)}\;,\\
&&U_2^\mu\xi_\mu =b^2\left (\frac{\delta\chi}{3}+\delta \eta_{lll}\right )+\frac{(\chi_\perp-\eta_{ll})v^2}{3}-\frac{a^2 (\chi+\lambda_\perp) [(\chi_\perp+\lambda_\perp)v^2 +b^2(\delta\lambda+\delta\chi)] }{4\varepsilon H^\parallel_\parallel(V,\xi)}\;,\\
&&H^\parallel_\parallel(V,\xi)=\frac{3\chi a^2+\lambda_\perp v^2+\delta\lambda\, b^2}{4\varepsilon}\;.
\eea
\eml
Here, $\delta\eta_{lll\perp}=4\eta_l-3\eta_{ll}-\eta_\perp$, $\delta\eta_{lll}=\eta_{ll}-\eta_l$, $\delta \chi= \chi_l- \chi_\perp$, and $\delta \lambda= \lambda_l- \lambda_\perp$. 
Because $\xi_\mu$ needs to be spacelike according to Eq.\ \eqref{C2}, no root with $v=0$ is allowed. Note that $v=0$ yields the following result: 
\[
	\xi_\mu \xi^\mu=(-u^\mu u^\nu+\Delta^{\mu\nu})\xi_\mu \xi_\nu=-a^2+v^2=-a^2<0\;,
\]
which clearly violates causality according to Eq.\ \eqref{C2}.
Upon inspection of Eqs.\ \eqref{matter1} and \eqref{matter2}, it is apparent that if either $\lambda_\perp$ or $\chi$, which give rise to the leading-order power in $a$, are zero, causality violating roots with $b=v=0$ arise. Building upon this observation and the known isotropic conformal case \cite{Bemfica:2017wps}, we can justify the following choice to ensure causality, i.e., 
\begin{equation}\label{A1}
	\lambda_\perp > 0 \;,\qquad \chi > 0 \;,\qquad \varepsilon > 0\;.
\end{equation}
We can rewrite $A$ by writing $b=l_\mu v^\mu=v\cos\theta$ ($0\le\theta\le\pi$) using the Cauchy-Schwarz inequality, as both vectors are orthogonal to $u^\mu$. In this case, $\Delta^{\mu\nu}$ defines a real inner product among these vectors. This results in the rewriting of $A$, Eq.\ \eqref{matter1}, as follows:
\be
A=\lambda_\perp(a^2-\tau v^2)\;,
\ee  
where 
\be
\label{tau}
\tau=\frac{\eta_\perp-\delta\eta_{\perp l}\cos^2\theta}{\lambda_\perp}\;. 
\ee
When considering the constraints  \eqref{A1}, the roots of $A=0$ obey \eqref{C1} and \eqref{C2} if, and only if,\footnote{See for instance Ref.\  \citep{Bemfica:2020zjp}.} $0\le \tau< 1$.\footnote{The equality $\tau=1$ may be used if the particles are massless.} 
Thus, if the conditions \eqref{A1} hold, the roots of $A=0$ are causal if, and only if,
\bea\label{Condition_root_A}
&&\lambda_\perp>\max(\eta_l,\,\eta_\perp) \geq 0\;.\label{condition_lambda}
\eea
However, the previous inequality does not lead to a sufficient condition for $\lambda_\perp$ when comparing the isotropic limit of this constraint, i.e., $\lambda > \eta$, and comparing with the corresponding one derived in the conformal isotropic case \cite{Bemfica:2017wps}. Therefore, it is needed to analyze the roots of \eqref{matter2} as well. In general this analysis becomes cumbersome, and as is shown in Appendix \ref{app:specific-causaility}, it requires finding the roots of the following polynomial in $\varrho\equiv a^2/v^2$:
\be
\label{polynomial_p_rep]}
p(\varrho) = \sum_{i=0}^{3}\alpha_i \varrho^i\;.
\ee
This polynomial is found by writing Eq.\ \eqref{matter2} in the generic form given by Eq.\ \eqref{general_root}, with the coefficients $\alpha_i$ to be determined from the aforementioned equality. The causality condition is then stated in the statement \ref{st:general-case}:
Assuming $\alpha_3$ to be positive, causality then requires $p(\varrho)$ to be positive for $\varrho\geq 1$ and negative for $\varrho<0$, while satisfying the following inequality:    
\begin{equation}
	\label{general_cond_rep}
	18\alpha_0\alpha_1\alpha_2\alpha_3-4\alpha_2^3\alpha_0+\alpha_2^2\alpha_1^2-4\alpha_3\alpha_1^3-27\alpha_3^2\alpha_0^2\ge0\;.
\end{equation} 
We have explained the application of the generic conditions of statement \ref{st:general-case}. To simplify the analysis of the causality conditions, it is more instructive to consider specific cases that provide clearer results. From the constitutive relations \eqref{Definitions}, we realize that the spatial anisotropy appears explicitly in different dissipative fluxes of the energy-momentum tensor: in the scalar sector with $\delta\chi\neq 0$, in the vector sector with $\delta \lambda\neq 0$, or in the tensor sector through the differences between three different transport parameters, i.e, $\eta_{\perp}$, $\eta_{l}$, and $\eta_{ll}$. 
If the only source of anisotropy is the scalar sector, then, as stated in statement \ref{st:chi-causality}, causality requires the five inequalities given by Eqs. \eqref{tensor_1_conditions}, together with Eqs. \eqref{eq:onshell-positivity} and \eqref{condition_lambda}, to be satisfied. The aforementioned conditions \eqref{tensor_1_conditions} are, for example, satisfied by the following choices:
\begin{equation}\label{eq:chi-aniso-params}
	\lambda=\chi=10\eta\;,\qquad
	\chi_\perp=\delta\chi=15\eta/2\;,	
\end{equation}
with $\eta_{\perp}=\eta_{l}=\eta_{ll}$, and $\delta\lambda=0$. Taking the isotropic limit \eqref{L1}, the conditions \eqref{tensor_1_conditions} reduce to the ones of the conformal isotropic BDNK \cite{Bemfica:2017wps}, i.e.,  
\bml
\bea
\chi>4\eta\;,\qquad 
\lambda>\frac{3\chi\eta}{\chi-\eta}\;.
\eea
\eml
On the other hand, if the anisotropy is only in the vector sector, and the transport parameters of the tensor sector vanish, i.e., the system is shearless, as stated in statement \ref{st:shearless}, the condition \eqref{A1} is sufficient for causality, without any further constraint on $\lambda_l$.  

To check for causality in more general cases, one can use statement \ref{st:general-case}. For instance, in the example worked out entirely in Appendix \ref{app:causal-example} one finds the following conditions that respect causality:
\be 
\label{example}
\eta_\perp=\eta\,,\quad\eta_l=\frac{2\eta}{3}\,,\quad\eta_{ll}=\frac{5\eta}{6}\,,\quad\lambda_\perp=\frac{13\eta}{2}\,,\quad\lambda_l=6\eta\,,\quad\chi=5\eta\,,\quad\chi_\perp=\frac{11\eta}{2}\,,\quad\chi_l=\frac{16\eta}{3}\,,\qq{with}\eta>0\;.
\ee
The proof of linear stability for this choice of parameters may be found in Appendix\ \ref{app:stability-example}.

\section{Linear stability}
\label{sec:stability}

In this section, we study the stability of the linearized EOM for the hydrodynamic theory developed in Secs. \ref{sec:abdnk} and \ref{sec:entropy}. We adopt standard methods \cite{Hiscock_Lindblom_instability_1985,Bemfica:2019knx} and consider small perturbations of the hydrodynamic fields $\varepsilon$ and $u^\mu$ around a homogeneous background, which corresponds to a global equilibrium with constant hydrodynamic fields in flat spacetime. In particular, we assume the energy density to be $\varepsilon_0+\delta\varepsilon(t,x^i)$,  with $\varepsilon_0$ being constant, and the fluid's velocity to be  $u^\mu_0 +\delta u^\mu(t,x^i)$, with $u^\mu_0 \delta u_\mu = \order{\delta^2}$. The fluid velocity in equilibrium is $u^\mu_0 = \gamma\left(1,v^i\right)$, with\
$v^i$ being the components of the 3-velocity $\vb{v}$, and $\gamma=1/\sqrt{1-\vb{v}^2}$ the Lorentz factor. We then expand the equations of motion up to the first order in perturbations, which are assumed to be in the form of plane waves, i.e., $\delta\varepsilon(t,x^i)\to e^{-iT_0 x^\mu k_\mu}\delta\varepsilon(k^\mu)$ and 
$\delta u(t,x^i) \to e^{-i T_0 x^\mu k_\mu}\delta u(k^\mu)$
, where $k^\mu=(i\Gamma,k^i)$ and the presence of the equilibrium temperature $T_0$ in the exponent makes the modes $k^\mu$ dimensionless. Nontrivial solutions to the EOM may lead to imaginary solutions $\Gamma=\Gamma(k^i)$, where linear stability is verified if, and only if, $\Re(\Gamma)\le 0$.

The roots $\Gamma=\Gamma(k^i)$ that come from the EOM are usually cumbersome, so the analysis becomes very difficult. However, new results show that causality + linear stability in the local rest frame (LRF) leads to linear stability in any boosted frame. This relation has been shown to be true for strongly hyperbolic systems of equations, as demonstrated in Ref.\ \citep{Bemfica:2020zjp}. A recent study by Gavassino \cite{Gavassino:2021owo} demonstrated the validity of this relation for the general case. The author showed that if a mode grows for an observer A, it can be thought of as a parametrization of the time coordinate in that frame. If the theory is causal, this growth should be preserved between frames, otherwise there would be an inversion in the time direction. Accordingly, and since we already have the conditions for causality, we shall study linear stability in the LRF and, in some specific cases, linear stability in a homogeneous boosted frame (where $k^i=0$ but $v^i\ne 0$).

We begin by studying linear stability in the LRF, i.e., where $u^\mu_0=(1,0,0,0)$ and, as a consequence, $\delta u^0=0$. Furthermore, since $l$ is orthogonal to $u$, it can only have spatial components in the LRF, i.e., $l^\mu_0 = \left(0,l^i\right)$, with keeping $l^\mu l_\mu = 1$ in mind. We note that although $l^\mu$ must also be perturbed in order to preserve the aforementioned orthogonality, its perturbation does not contribute to the dissipative fluxes of \eqref{Definitions} up to the first order in perturbations.

Let us define $\delta\bar{\varepsilon}=\delta\varepsilon/(\varepsilon+P)$ and $\delta\bar{T}^{\mu\nu}=\delta T^{\mu\nu}/(\varepsilon+P)$, together with the dimensionless quantities $\bar{\eta}_l=\eta_l/s$, $\bar{\eta}_{ll}=\eta_{ll}/s$, $\bar{\eta}_\perp=\eta_\perp/s$, $\bar{\lambda}_l=\lambda_l/s$, $\bar{\lambda}_\perp=\lambda_\perp/s$, $\bar{\chi}=\chi/s$, $\bar{\chi}_l=\chi_l/s$, and $\bar{\chi}_\perp=\chi_\perp/s$, where $s=(\varepsilon+P)/T=4\varepsilon/3T$ is the equilibrium entropy density. We may also define $\bar{k}=l_i k^i$, $\kappa^i=\Xi^i_jk^j=k^i-l^i\bar{k}$, and $\kappa=\sqrt{\kappa_i\kappa^i}$ in order to perform the decomposition \[\delta u^i=l^i\delta u_L+\frac{\kappa^i}{\kappa} \delta u_\parallel+\delta u^i_\perp,\] where $\delta u_L=l_i\delta u^i$, $\delta u_\parallel=\kappa_i\delta u^i/\kappa$, and $\delta u_\perp^i=\Xi^i_j\delta u^j-\kappa^i\delta u_\parallel$. With that in mind, we obtain the following equations for the modes:    
\bml
\label{modes}
\bea
&&\partial_\nu \delta\bar{T}^{0\nu}=M_{11}\delta\bar{\varepsilon}+M_{12}\delta u_L+M_{13}\delta u_\parallel = 0\;,\label{modes_a}\\
&&l_i\partial_\nu\delta\bar{T}^{i\nu}= M_{21}\delta\bar{\varepsilon}+M_{22}\delta u_L+M_{23}\delta u_\parallel = 0\;,\\
&&\frac{\kappa_i}{\kappa}\partial_\nu\delta\bar{T}^{i\nu}= M_{31}\delta\bar{\varepsilon}+M_{32}\delta u_L+M_{33}\delta u_\parallel = 0\;,\label{mode_c}\\
&&\omega^i_j\partial_\nu\delta\bar{T}^{j\nu}= \left [\bar{\lambda}_\perp\Gamma^2+\Gamma+\bar{\eta}_\perp k^2+(\bar{\eta}_l-\bar{\eta}_\perp)\bar{k}^2\right ]\delta u^i_\perp=0\;,\label{modes_d}
\eea
\eml
where $\omega^{ij}=\Xi^{ij}-\kappa^i\kappa^j/\kappa^2$ is the projector orthogonal to $\kappa^i$ and $l^i$,  $k^2=k_ik^i$, and
\bml
\bea
M_{11}&=&\bar{\chi}  \Gamma^2+\Gamma-\frac{k^2 \bar{\lambda}_\perp}{3}+\frac{\bar{k}^2}{3}(\bar{\lambda}_\perp -\bar{\lambda}_l)\;,\\
M_{12}&=&i \bar{k} \left [\left(\bar{\lambda}_l+\bar{\chi}\right )  \Gamma+1\right]\;,\\
M_{13}&=&i \kappa \left[\left (\bar{\lambda}_\perp+\bar{\chi}\right )\Gamma+1\right]\;,\\
M_{21}&=&\frac{i}{3} \bar{k} \left[\left (\bar{\lambda}_l+ \bar{\chi}_l\right )\Gamma+1\right]\;,\\
M_{22}&=&\bar{\lambda}_l\Gamma^2 +\Gamma+\kappa^2 \bar{\eta}_l+\frac{(4
	\bar{\eta}_{ll}-\bar{\chi}_l)\bar{k}^2}{3}\;,\\
M_{23}&=&\frac{\bar{k}\kappa}{3} \left(3 \bar{\eta}_l-\bar{\chi}_l-2\bar{\eta}_{ll}\right)\;,\\
M_{31}&=&\frac{i\kappa}{3}\left[\left (\bar{\chi}_l+\bar{\lambda}_\perp\right )\Gamma+1\right]\;,\\
M_{32}&=&\frac{ \bar{k}\kappa}{3} \left(3 \bar{\eta}_l-\bar{\chi}_l-2 \bar{\eta}_{ll}\right)\;,\\
M_{33}&=&\Gamma^2 \bar{\lambda}_\perp+\Gamma+\frac{k^2}{3} \left(-\bar{\chi}_l+\bar{\eta}_{ll}+3 \bar{\eta} _\perp\right)
+\frac{\bar{k}}{3} \left(3 \bar{\eta}_l+\bar{\chi}_l-\bar{\eta}_{ll}-3 \bar{\eta}_\perp\right)\;.    
\eea
\eml
Note that the transverse modes $\delta u_\perp^i$ decouple from the rest and give the shear channel polynomial equation
\be
\label{shear}
\bar{\lambda}_\perp\Gamma^2+\Gamma+\bar{\eta}_\perp k^2+(\bar{\eta}_l-\bar{\eta}_\perp)\bar{k}^2=0\;.
\ee
The equations for the longitudinal modes \eqref{modes_a}--\eqref{mode_c} are nontrivial, i.e., give nonzero solutions for the perturbations, when
\be
\label{sound}
\det(M)=0\;,
\ee
where $M=[M_{ij}]_{3\times 3}$. The roots of Eq.\ \eqref{sound} give the wave modes of the sound channel. It is worth mentioning that the mode equations in a Lorentz boosted frame are obtained by performing a boost which yields to the following change 
\begin{align}\label{B1}
	  & \bar{k}\to l_\mu k^\mu\,,\quad k^2\to-\gamma^2(\Gamma+ik_iv^i)^2+\Gamma^2+k^ik_i\,, 
	\quad 
	\kappa^2\to k^2	-(l_\mu k^\mu)^2\;,\quad\Gamma\to\gamma(\Gamma+ik_iv^i)\;.
\end{align}
In the boosted frame, $u^\mu=\gamma(1,v^i)$ and $l^\mu=(\tilde{l}^kv_k,\tilde{l}^i)/\sqrt{1-v^l \tilde{l}_l}$, with $\tilde{l}^i$ being a unitary 3-vector that coincides with the anisotropic unitary 3-vector $l^i$ in the LRF ($v=0$). 

In the LRF case, by means of the Cauchy-Schwarz inequality, one may set $l_ik^i=k x$, where $-1\le x\le 1$ covers all possible directions of $k^i$. Then, Eq.\\ \eqref{shear} can be written as
\be
\label{shear2}
\bar{\lambda}_\perp\Gamma^2+\Gamma+\left [\bar{\eta}_\perp\left(1-x^2\right)+\bar{\eta}_l \right ]k^2=0.
\ee 
The roots of the above equation have zero or negative real parts if, and only if, $\bar{\lambda}>0$ and $\bar{\eta}_\perp+x^2(\bar{\eta}_l-\bar{\eta}_\perp)\ge0$. Both conditions are guaranteed by the constraints  \eqref{A1} and \eqref{Condition_root_A} because $1-x^2 \ge 0$, and we have that either $\bar{\eta}_\perp > \bar{\eta} \ge 0 $ or $\bar{\eta} >\bar{\eta}_\perp  \ge 0$. Hence, Eqs. \eqref{A1} and \eqref{Condition_root_A} lead to the linear stability of the shear channel in the LRF.

Now, let us consider \eqref{shear2} in the boosted homogeneous case, i.e., with vanishing $k^i$. By employing again the Cauchy-Schwarz inequality, we can write $(l_\mu k^\mu)^2=(\Delta^{\mu\nu}k_\mu k_\nu) y^2$, with $-1 \leq y \leq 1$, and then apply \eqref{B1} to obtain 
\be
\Gamma\left [\left(\bar{\lambda}_\perp -(1-y^2)v^2\bar{\eta}_\perp -  y^2 v^2\bar{\eta}_l\right) \gamma\Gamma+1\right ]=0\;,
\ee 
which has roots for all $v^2\in[0,1)$ with nonpositive real parts if
	$\bar{\lambda}_\perp \ge \bar{\eta}_\perp\left(1-y^2\right)+\bar{\eta}_ly^2$ and $\bar{\lambda}_\perp>0$.  Both inequalities are guaranteed by the causality condition \eqref{Condition_root_A} and Eq.\ \eqref{A1}, according to which $\bar{\eta}_\perp\left(1-y^2\right)+\bar{\eta}_ly^2 \leq \max(\bar{\eta}_l,\bar{\eta}_\perp) <  \bar{\lambda}_\perp$. In this condition we clearly see the connection between stability in any frame and causality+stability in the LRF \cite{Gavassino:2021owo}.
	
	The polynomial for the sound channel in \eqref{sound} is of power six, with complicated coefficients that depend on all transport parameters as well as $\bar{k}$ and $k^2$. The analysis of this type of polynomials is extremely complex, so it is more convenient and better to examine each set of parameters separately. As an example, the stability of the causal set of parameters \eqref{example} is demonstrated in Appendix \ref{app:stability-example}.
	\section{Bjorken flow}
	\label{sec:bjorken}
	In this section, we apply the formalism developed in this work to the case of Bjorken flow, i.e., when the fluid's velocity is $u^\mu= \left(1,0,0,0,\right)$ in the so-called Milne coordinates $\left(\tau,x,y,\xi\right)$, which are related to the usual Minkowski coordinates via $\tau=\sqrt{t^2-z^2}$ and $\xi = \half \log[(t+z)/(t-z)]$. From a mathematical perspective, one can choose the spacelike anisotropic vector to be in any of the $x$, $y$, and $\eta$ directions. However, the physical picture of the heavy ion collisions suggests 
	\begin{equation}\label{eq:bjorken-l}
		l^\mu = \inv{\tau} \left(0,0,0,1\right)\,,	
	\end{equation}
	to be the right choice.  Alternatively, other choices may be possible, but one should be cautious of potential nonphysical issues when making such a choice, as illustrated in Appendix \ref{app:wrong}. By equating $u$ and $l$ into Eqs. \eqref{Definitions} and using the Bjorken symmetries, the dissipative fluxes  reduce to 
	\begin{subequations}
		\bea
		\mathcal{E}^{(1)}&=&\tilde{\chi}^3T^3\left(\inv{\tau}+\frac{3\dot{T}}{T}\right)\,,
		\\
		\mathcal{P}_l^{(1)}&=&T^3\left(\frac{\tilde{\chi}_l-4\tilde{\eta}_{ll}}{3\tau}+\frac{3\tilde{\chi}_l\dot{T}}{T}\right)\,,
		\\
		\mathcal{P}_\perp^{(1)}&=&T^3\left(\frac{\tilde{\chi}_\perp+2\tilde{\eta}_{ll}}{3\tau}+\frac{3\tilde{\chi}_\perp\dot{T}}{T}\right)\,,
		\eea
	\end{subequations}
	and 
	\begin{equation}
		\pi_\perp^{\mu\nu}=0\,,\qquad
		W_{\perp l}^\mu=0\,,\qquad
		W_{\perp u}^\mu=0\,,\qquad
		M=0\,.
	\end{equation}
	Namely, anisotropy only appears in the scalar sector with three independent relevant transport parameters, $\chi = \tilde{\chi} T^3$,  $\chi_\perp = \tilde{\chi}_\perp T^3$, and $\eta_{ll}= \tilde{\eta}_{ll}  T^3$. Recall that in the isotropic case, the relevant transport parameters for the Bjorken flow are shear viscosity $\eta$, and $\chi$ \cite{Bemfica:2017wps,Shokri:2020cxa}. Similar to the cases of IS \cite{Heller:2015dha} and isotropic conformal BDNK, the fluid's evolution is governed by only one equation, 
	\begin{eqnarray}\label{eq:bjorken-eom}
		&&9\tilde{\chi}\frac{\tau^2\ddot{T}}{T} + 18\tilde{\chi}\frac{\tau^2\dot{T}^2}{T^2}+ \left(\frac{3\tau(9\tilde{\chi}-\tilde{\chi}_\perp)}{T}+12\tau^2\right)\dot{T}+4\tau T +3\tilde{\chi}- 2\tilde{\chi}_\perp -  4\tilde{\eta}_{ll} = 0\,.
	\end{eqnarray}
	At late times, the solution to the previous equation can be written as a power series, \begin{equation}\label{eq:late-time-tau}
	T = \frac{\Lambda}{(\Lambda\tau)^{1/3}}\left(1-\frac{\tilde{\eta}_{ll}}{2(\Lambda\tau)^{2/3}}-\frac{\tilde{\eta}_{ll}(\tilde{\chi}_l+5\tilde{\chi}_\perp)}{24(\Lambda\tau)^{2/3}}+\cdots\right)\,,
	\end{equation}
	where $\Lambda$ is a constant with energy dimensions. We clarify that when we refer to late times, we mean  $\Lambda\tau \gg 1$. One may assume that up to the first order, the power series solution is equal to the one of the isotropic conformal case, to reduce the number of free parameters.  Such an assumption gives rise to $\eta_{ll} = \eta$.
	
	To gain more insight into the physical implications of \eqref{eq:bjorken-eom}, especially in the far from equilibrium regime, we assume the dimensionless parameters $w=T\tau$ \footnote{The variable $w$ is proportional to the inverse Knudsen number for the conformal case.} and $f(w)=\frac{\tau}{w}\dv{w}{\tau}$ \cite{Heller:2015dha}. The EOM \eqref{eq:bjorken-eom} reduces to a first-order nonlinear differential equation, 
	\begin{eqnarray}\label{eq:bjorken-feom}
		\frac{9\tilde{\chi}}{4}f(w)^2+
		w f(w)\left(1+\frac{3}{4}\tilde{\chi}f'(w)\right)
		- \frac{6\tilde{\chi}+\tilde{\chi}_\perp}{2}f(w) +\frac{3\tilde{\chi}+\tilde{\chi}_\perp-\tilde{\eta}_{ll}}{3}-\frac{2w}{3} = 0\,.
	\end{eqnarray} 
	The late-time expansion \eqref{eq:late-time-tau} is written in terms of the variable $w$ as 
	\begin{equation}\label{eq:late-time-f}
		f(w) = \frac{2}{3} + \frac{\tilde{\eta}_{ll}}{3w}+\frac{\tilde{\eta}_{ll(\tilde{\chi}_l+5\tilde{\chi}_\perp)}}{18w^2} + \order{\inv{w^3}}\,,
	\end{equation}
 which is valid when $w\gg 1$.
	The pressure anisotropy, $\mcA = \left(\mcP_\perp- \mcP_l\right)/P$, in the isotropic conformal BDNK, and in contrast to IS theory, is purely determined by shear viscosity and does not depend on $f$. In the anisotropic case, the situation is different because $\mcA$ receives contribution from $f(w)$,
	\begin{equation}
		\mcA = \frac{2}{3}\frac{\tilde{\chi}_\perp-\tilde{\chi}_l}{w} \left(f-\frac{2}{3}\right) + \frac{6\tilde{\eta}_{ll}}{w}\,.
	\end{equation}
	Note that at late times $f(w) \sim 2/3 + \order{w^{-1}}$ so both isotropic and anisotropic first-order BDNK theories share the same forward attractor. This is expected since asymptotically the system relaxes towards the thermal equilibrium. Following Ref.\ \cite{Heller:2015dha}, we assume a correction to the first-order on-shell terms in \eqref{eq:late-time-f}, 
	\begin{equation}
		f(w) = \frac{2}{3} + \frac{\tilde{\eta}_{ll}}{3w}+\delta f(w)\,.
	\end{equation}
	By equating the previous expression into Eq.\ \eqref{eq:bjorken-feom} while assuming $\delta f \ll f$, and expanding in $1/w$  around $w\to \infty$, we obtain 
	\begin{equation}
 \label{eq:pert}
		\delta f(w) \sim \exp(-\frac{2w}{\tilde{\chi}}) w^{\frac{\tilde{\eta}_{ll}+\tilde{\chi}_\perp}{\tilde{\chi}}}\,.
	\end{equation}
Since $\chi>0$ due to causality, at late times the perturbation \eqref{eq:pert} decays faster than the perturbative terms of the late-time expansion. The relation between the coefficients in the above form and the analytical structure of the Borel resummation might be of interest, but it will not be discussed here.

	Equation \eqref{eq:bjorken-feom} can also be studied for the existence of pullback attractors. Following the transasymptotic and dynamical systems methods outlined in Refs. \cite{Behtash:2017wqg,Behtash:2018moe,Behtash:2019txb,Behtash:2020vqk,Kamata:2022jrc,Kamata:2022ola} one can show that the initial value which rises to the attractor solution is found by expanding \eqref{eq:bjorken-feom} around $w=0$, 
	\begin{equation}
		f(w\ll 1) = \frac{7}{9} -\frac{\tilde{\chi}-\tilde{\chi}_\perp}{9\tilde{\chi}}+ \frac{\sqrt{12\tilde{\eta}_{ll}\tilde{\chi}+\tilde{\chi}_\perp^2}}{9\tilde{\chi}}\,.
	\end{equation}
	In the following we use the slow-roll approximation \cite{Heller:2015dha}. Namely, we assume $\abs{f'}$ to be much smaller than $\abs{f}$ and expand \eqref{eq:bjorken-feom} in terms of $f'/f$ to obtain an algebraic equation. This leads to two solutions for $f(w)$ that we expand around $w\to\infty$ and compare with \eqref{eq:late-time-f}, to recognize the stable one, 
	\begin{equation}\label{eq:slow-roll}
		f(w)_{\rm slow roll} = \frac{7}{9} -\frac{\tilde{\chi}-\tilde{\chi}_\perp}{9\tilde{\chi}} -\frac{2w}{9\tilde{\chi}}+\frac{\sqrt{\left(2w-\tilde{\chi}_\perp\right)^2+12\tilde{\eta}_{ll}\tilde{\chi}}}{9\tilde{\chi}}\,.
	\end{equation}
	If at early times, $f$ gets larger than 1 then the fluid experiences reheating.  To prevent reheating we must have
	\begin{equation}
		\chi_l > 4 \eta_{ll} >  0\,,
	\end{equation}
	which can be shown that is equivalent to the condition for stability and causality and reduces to $\chi > 4\eta$ in the isotropic limit. Therefore, causality forbids reheating for the Bjorken model discussed in this section.

	Finally, we might assume an off-shell, or nonphysical, entropy current $S^\mu_{\rm off}$, that is, a current determined from \eqref{Entropy_current} without considering the power counting discussed in Sec.\ \ref{sec:entropy}. If causal and stable parameters are chosen for the attractor solution \eqref{eq:slow-roll}, $\nabla\cdot S_{\rm off}$ will be negative at early times. However, if causality and stability are violated, the divergence will always be positive. $\nabla\cdot S_{\rm off}$ approaches zero rapidly and then changes sign in the stable and causal case. This seems to be a typical characteristic of BDNK theories, which may be related to the fact that the domain of applicability must be explicitly considered.
	\section{Conclusions}
	\label{sec:conclusions}
	In this work, we explored the constraints imposed by causality in anisotropic hydrodynamics. Our approach involved deriving the simplest and most general form of the first-order anisotropic energy-momentum tensor, based on the invariance under the little group $SO(2)$ and the second law of thermodynamics. We demonstrated that our theory is both causal and stable, and also well-posed whether or not it is coupled to gravity. Furthermore, we showed that the standard isotropic BDNK theory can be recovered as a limit of our novel approach.

We verify the linear stability and causality of our theory in both the boosted and local rest frames, confirming the generality of recent results obtained by Gavassino \cite{Gavassino:2021owo}. For the specific case of Bjorken flow, we investigate the causality conditions necessary to ensure the existence of forward and pullback attractors. Our findings reveal that the behavior of these attractors at early and late times is constrained by causality conditions. Violation of these conditions results in a reheating effect, which is the increase of temperature from its initial value at very early times.

There are several potential avenues for future research that can build on our work. First, it would be valuable to investigate how our novel anisotropic theory can be derived from a coarse-graining approach, as is typically done in relativistic kinetic theory.  Additionally, extending our analysis to the most general nonconformal case for both charged and uncharged fluids, with and without gravity, could have significant implications in other fields such as cosmology \cite{Bemfica:2022dnk}. We are extremely confident that valuable insights will emerge with  these investigations. For instance, these works will provide the theoretical basis and guidance for a meticulous exploration of potential causality violations in numerical simulations of anisotropic hydrodynamics for heavy ion collisions, and thus, complementing recent phenomenological analysis carried out by various research groups \cite{Plumberg:2021bme, Chiu:2021muk}. Our work may also impact the development of a BDNK-type theory for resistive dissipative magnetohydrodynamics \cite{Denicol:2019iyh,Armas:2022wvb,Dash:2022xkz,Gavassino:2023qnw}, where the magnetic field serves a similar purpose to the anisotropy vector $l^\mu$. 

Finally, while our approach has been successful in analyzing the near-equilibrium regime, it would be valuable to extend our techniques to the extreme far-from-equilibrium dynamics. This could provide valuable insights into how causality affects the spacetime evolution of these systems. These fascinating research problems are important and require further investigation in the future.

 \begin{acknowledgments}
 M. S. was supported by the Deutsche Forschungsgemeinschaft (DFG, German Research Foundation) through the Collaborative
Research Center CRC-TR 211 ``Strong-interaction matter under extreme conditions'' - Project No.\ 315477589 -
TRR 211 and by the State of Hesse within the Research Cluster ELEMENTS (Project ID 500/10.006). M. M. was supported in part by the U.\S.\ Department of Energy Grant No. DE-FG02-03ER41260 and BEST (Beam Energy Scan Theory) DOE Topical Collaboration. 

\end{acknowledgments}
	
	\appendix
	\section{\MakeUppercase{Anisotropic free energy density}}
	\label{Appendix_A}
	In this appendix, we present the thermodynamic relations of an anisotropic fluid. To this end, let us assume the leading-order generating functional of the fluid to be \cite{Jensen:2012jh,Banerjee:2012iz}
	\begin{equation}
		W_0 = \int \dd[4]{x}\sqrt{-g}\mcF(T,\ell)\,.
	\end{equation} 
	Here, $\mcF$ is the free energy that is to be determined and $\ell=\sqrt{g_{\mu\nu}l^\mu l^\nu}$, which is set to unity at the end.\footnote{$\ell$ is similar to the parameter $\xi$ for the superfluid case in \cite{Jensen:2012jh}. Note that, $\ell$ varies only because of the metric variation.} The energy-momentum tensor can then be derived from 
	\begin{equation}\label{eq:em-tensor-def}
		T^{\mu\nu} =
		\frac{2}{\sqrt{-g}} \fdv{W}{g_{\mu\nu}}=\Bigg(\mcF\delta\sqrt{-g}+\sqrt{-g}\left(\pd{\mcF}{T}\right)_{\ell}\delta T+\sqrt{-g}\left(\pd{\mcF}{\ell}\right)_{T}\delta\ell\Bigg)\,.
	\end{equation}
	The variation of temperature and $\ell$ are 
	\begin{equation*}
		\delta T = \half T u^\mu u^\nu \delta g_{\mu\nu}\,,\qquad 
		\delta\ell = \half l^\mu l^\nu \delta g_{\mu\nu}\,. 
	\end{equation*} 
	Plugging the above relations into \eqref{eq:em-tensor-def}, we obtain 
	\begin{equation}\label{eq:em-tensor-2}
		T^{\mu\nu}_{0}=\left(-\mcF+T\left(\pd{\mcF}{T}\right)\at[\Big]{\ell}\right)u^\mu u^\nu+\left(\pd{\mcF}{\ell}\right)\at[\Big]{T}l^\mu l^\nu+\mcF \Delta^{\mu\nu}\,.
	\end{equation}
	The pressure and longitudinal pressure are defined as
	\begin{align*}
		  & P \equiv \inv{3}\Delta_{\mu\nu} T^{\mu\nu} = \mcF + \inv{3}\left(\pd{\mcF}{\ell}\right)_{T}\,, 
		\qquad
		P_l \equiv l_\mu l_\nu T^{\mu\nu} = \mcF + \left(\pd{\mcF}{\ell}\right)_{T}\,.
	\end{align*}
	Plugging \eqref{eq:em-tensor-2} into the above, we recognize the free energy,
	\begin{equation}
		\mcF = \frac{3P-P_l}{2} \equiv P_\perp\,.		
	\end{equation}
	After identifying the free energy with the transverse pressure, one can observe the relations between thermodynamic quantities,
	\begin{align}\label{eq:aniso-euler}
		  & \ep \equiv u_{\mu}u_\nu T^{\mu\nu} = -P_\perp+T\left(\pd{P_\perp}{T}\right)_{\ell}\,,\qquad 
		\left(\pd{P_\perp}{\ell}\right)_{T} = P_l - P_\perp\,.
	\end{align}
	The first equation above is Euler's equation for the anisotropic fluid that determines its entropy density,
	\begin{equation}\label{eq:entropy-density}
		s=\frac{\ep+\mcF}{T}=\left(\pd{P_\perp}{T}\right)_{\ell}\,.
	\end{equation}
	From the above definition of the entropy density and \eqref{eq:aniso-euler}, we find
	\begin{eqnarray*}
		\dd{\ep} &=& - \left(P_l - P_\perp\right)\dd{\ell} + T\dd{s}\,\qquad
		\dd{P_\perp} = \left(P_l - P_\perp\right)\dd{\ell} + s\dd{T}\,. 
	\end{eqnarray*}
	At this point, we set $\ell=1$ to find the Gibbs-Duhem relation and the first law of thermodynamics for the anisotropic fluid
	\begin{eqnarray}\label{eq:ah-thermo-laws}
		\dd{\ep} &=&  T\dd{s}\,\qquad
		\dd{P_\perp} = s\dd{T}\,. 
	\end{eqnarray}
	Finally, the entropy current is found from the covariant form of \eqref{eq:entropy-density}
	\begin{equation}
		T s_\mu = P_\perp u_\mu - u^\nu T_{\mu\nu}\,.
	\end{equation} 
	\section{\MakeUppercase{Details of causality analysis}}\label{app:causality}
	In this appendix we give the mathematical details of causality analysis that are not presented in Sec.\ \ref{sec:causality}.
	The system of equations arising from \eqref{Projected_EOM} and \eqref{GR_eq} can be written as \footnote{One may notice that causality is blind to the leading-order anisotropy of pressure since it does not appear in the principal part of the system of PDEs.}
	\bml
	\label{EOM4}
	\bea
	&&\frac{3\chi u^\alpha u^\beta+\lambda_\perp \Delta^{\alpha\beta}+\delta\lambda\, l^{\alpha}l^{\beta}}{4\varepsilon}\partial_\alpha\partial_\beta \varepsilon
	+\left [(\chi+\lambda_\perp) u^{(\alpha}\delta^{\beta)}_\nu+\delta\lambda\, l^{(\alpha} u^{\beta)} l_\nu \right ] \partial_\alpha\partial_\beta u^\nu\nonumber\\
	&&\hspace{4cm}+h_{a}^{\parallel,\,\alpha\beta}(\varepsilon,u,g)\partial_\alpha\partial_\beta g_a=b^\parallel(\partial\varepsilon,\partial u,\partial g)\;,\\
	&&\frac{(\chi_\perp+\lambda_\perp) \Delta^{\mu(\alpha} u^{\beta)}+(\delta\lambda+\delta\chi)\, l^\mu l^{(\alpha} u^{\beta)}}{4\varepsilon}\partial_\alpha\partial_\beta\varepsilon+C^{\mu\alpha\beta}_\nu\partial_\alpha\partial_\beta u^\nu+h_a^{\mu,\,\alpha\beta}(\varepsilon, u,g)\partial_\alpha\partial_\beta g_a\nonumber\\
	&&\hspace{4cm}=b^{\mu}(\partial\varepsilon,\partial u, \partial g)\;,\\
	&&g^{\alpha\beta}\partial_\alpha\partial_\beta g_a= b_{a}(\partial\varepsilon,\partial u,\partial g)\;,
	\eea
	\eml
	where there is an implicit sum over repeated $a,b=\mu\nu$, with $\mu\le \nu$, over the values $00,\,01,\,02,\,03,\,11,\,12,\,13,\,22,\,23$, and $33$. Furthermore,
	$\delta\lambda=\lambda_l-\lambda_\perp$, $\delta\chi=\chi_l-\chi_\perp$, and we have defined 
	\bea
	&&C^{\mu\alpha\beta}_\nu\equiv \delta\eta_{lll\perp}\,l^\mu l^\beta l^{\alpha}l_\nu+\left (\frac{\delta\chi}{3}+\delta \eta_{lll}\right )\,l^\mu l^{(\beta}\delta^{\alpha)}_\nu+\delta\eta_{lll}\,\Delta^{\mu(\beta}l^{\alpha)} l_\nu +\frac{(\chi_\perp-\eta_{ll})\Delta^{\mu(\beta}\delta^{\alpha)}_\nu}{3} \nonumber\\
	&&\hspace{2cm}+\delta\lambda\, l^\mu l_\nu u^\alpha u^\beta +\delta\eta_{\perp l}\Delta^{\alpha\beta}l^\mu l_\nu+\left [\lambda_\perp u^\alpha u^\beta-\eta_\perp\Delta^{\alpha\beta}+\delta\eta_{\perp l} l^\alpha l^\beta\right ]\delta^{\mu}_\nu\;,
	\eea
	wherein $\delta\eta_{lll\perp}=4\eta_l-3\eta_{ll}-\eta_\perp$, $\delta\eta_{lll}=\eta_{ll}-\eta_l$ and $\delta \eta_{\perp l}=\eta_\perp-\eta_l$. 
	We may rewrite \eqref{EOM4} in the matrix form 
	\be
	\label{Matrix_EOM}
	H^I_J(V,\partial) V^J+b^I=0,
	\ee
	where $I,J$ take the values $\parallel$, $\mu=0,1,2,3$, and $a=00,01,\cdots,33$, $V^I=(\varepsilon, u^\nu,g_b)$, 
	\[H(V,\partial)^I_J=h(V)^{I,\,\alpha\beta}_J\partial_\alpha\partial_\beta\] is a $15\times 15$ matrix linear operator, and 
	\bml
	\bea\label{eq:relevant-hab}
	h^{\parallel,\,\alpha\beta}_\parallel&=&\frac{3\chi u^\alpha u^\beta+\lambda_\perp \Delta^{\alpha\beta}+\delta\lambda\, l^{\alpha}l^{\beta}}{4\varepsilon},\\
	h^{\parallel,\,\alpha\beta}_\nu&=&(\chi+\lambda_\perp) u^{(\alpha}\delta^{\beta)}_\nu+\delta\lambda\, l^{(\alpha} u^{\beta)} l_\nu ,\\
	h^{\mu,\,\alpha\beta}_\parallel&=&\frac{(\chi_\perp+\lambda_\perp) \Delta^{\mu(\alpha} u^{\beta)}+(\delta\lambda+\delta\chi)\, l^\mu l^{(\alpha} u^{\beta)}}{4\varepsilon},\\
	h^{\mu,\,\alpha\beta}_\nu&=&C^{\mu\alpha\beta}_\nu,\\
	h^{a,\,\alpha\beta}_b&=&\delta^a_b g^{\alpha\beta},\text{ where }\delta^a_b=\delta^\mu_\lambda \delta^\nu_\sigma \text{  when }a=\mu\nu\text{ and }b=\lambda\sigma,\\
	h^{a,\,\alpha\beta}_\parallel&=&h^{a,\,\alpha\beta}_\mu=0.
	\eea
	\eml
	The remaining expressions $h^{\mu,\,\alpha\beta}_b(V)$ and $b^I(\partial V)$ are irrelevant to what follows. The principal part of each equation $I$ is contained in $H(V,\partial)^I_J V^J$. It is worth mentioning that all terms $b^I(\partial V)$ are functions of at most the first-order derivative of the variables $V^I=(\varepsilon,u^\nu,g_a)$, with products containing first-order derivatives such as $(\partial V^K)^n$, $(\partial V^K)^n(\partial V^L)^m$, among others, being allowed. On the other hand, $h^{I,\,\alpha\beta}_J(V)$ are functions of the variables $V^I$ only and not their derivatives. Therefore, this is a quasilinear PDE system and the usual tools to compute causality apply. The characteristic surfaces $\{\Phi(x)=0\}$ are determined by the principal part of the equations by solving the characteristic equation $\det[H(V^k,\xi)]=0$, with $\xi_\mu=\nabla_\mu \Phi$ \cite{Courant_and_Hilbert_book_2,ChoquetBruhatGRBook}. Note that the components of the matrix $H(V,\xi)$ are $H^I_J(V,\xi)=h(V)_J^{I,\alpha\beta}\xi_\alpha\xi_\beta$. 
	The system is causal if, for any given real $\xi_i$, (C1) the roots $\xi_0=\xi_0(\xi_i)$ of the characteristic equation are real and (C2) $\xi_\alpha=(\xi_0(\xi_i),\xi_i)$ is spacelike or lightlike, i.e., $\xi_\mu \xi^\mu\ge0$. Condition (C2) guarantees that the surfaces $\{\Phi(x)=0\}$ are timelike or lightlike, ensuring that there is no superluminal information. \footnote{Note that the matrix $H(V^ k,\xi)$ is invertible only when $\xi$ is timelike, i.e., solutions are only possible over spacelike or lightlike hypersurfaces $\Phi$.} 
		
	We may now compute the characteristic equation, for which we must compute the determinant of the matrix $H(V,\xi)$ that reads
	\bea
	\label{h}
	\det[H(V,\xi)]&=&\det\bbm
	H^\parallel_\parallel(V,\xi) & H^\parallel_\nu(V,\xi) & H^\parallel_b(V,\xi)\\
	H^\mu_\parallel(V,\xi) & H_\nu^{\mu}(V,\xi) & H_b^{\mu}(V,\xi)\\
	0_{10\times 1} & 0_{10\times 4} & \xi_\mu\xi^\mu I_{10} \ebm \nonumber\\
	&=&(\xi_\mu\xi^\mu)^{10}M\;,
	\eea
	where 
	\be
	\label{M1}
	M=\det\bbm
	H^\parallel_\parallel(V,\xi) & H^\parallel_\nu(V,\xi) \\
	H^\mu_\parallel(V,\xi) & H_\nu^{\mu}(V,\xi)
	\ebm,
	\ee
	wherein $I_{10}$ is the $10\times 10$ identity matrix and $0_{m\times n}$ is the $m\times n$ null matrix. 
	Out of 30 roots of the characteristic equation, 20 are the real roots $\xi_\mu\xi^\mu=0$ coming from the pure gravity sector and, thus, are causal. These are expected to be lightlike roots since pure gravity fields have no mass. If matter instead of pure radiation, which is permitted for a conformal fluid, is treated, the remaining roots are expected to be spacelike.
	As for the matter sector, let us define $a=u^\mu\xi_\mu$, $b=l^\mu \xi_\mu$, $v^\mu=\Delta^{\mu\nu}\xi_\mu$, and $v=\sqrt{\xi_\mu\xi_\nu\Delta^{\mu\nu}}$.
	Then, from \eqref{M1} one obtains that
	\bea
	\label{M}
	&&M=H^\parallel_\parallel(V,\xi)\det\bbm
	H_\nu^{\mu}(V,\xi)-\frac{H^\mu_\parallel(V,\xi) H^\parallel_\nu(V,\xi)}{H^\parallel_\parallel(V,\xi)} \nonumber\\
	\ebm\nonumber\\
	&&=H^\parallel_\parallel(V,\xi)\det\left[A\delta^{\mu}_\nu+U_1^\mu l_\nu +U_2^\mu \xi_\nu \right]\nonumber\\
	&&=A^2 H^\parallel_\parallel(V,\xi)\Big[A^2+A\left (U_1^\mu l_{\mu}+U_2^\mu \xi_{\mu}\right )+U_1^\mu l_{\mu}U_2^\nu \xi_{\nu}-U_1^\mu \xi_{\mu}U_2^\nu l_{\nu}
	\Big]\;,
	\eea   
	where
	\bml
	\bea
	A&=&\lambda_\perp a^2-\eta_\perp v^2+\delta\eta_{\perp l} b^2\;,\\
	U_1^\mu&=&b^2\delta\eta_{lll\perp}\,l^\mu +a^2\delta\lambda\,l^\mu+\delta\eta_{\perp l} v^2 l^\mu+b\delta\eta_{lll}\,v^\mu-\frac{ab\delta\lambda\, H^\mu_\parallel(V,\xi) }{H^\parallel_\parallel(V,\xi)}\;,\\
	U_2^\mu&=&b\left (\frac{\delta\chi}{3}+\delta \eta_{lll}\right )\,l^\mu+\frac{(\chi_\perp-\eta_{ll})v^\mu}{3}-\frac{a (\chi+\lambda_\perp) H^\mu_\parallel(V,\xi) }{H^\parallel_\parallel(V,\xi)}\;,
	\eea
	\eml
	In particular, let us write explicitly
	\bml
	\bea
	&&U_1^\mu l_\mu =b^2\delta\eta_{lll\perp} +a^2\delta\lambda+\delta\eta_{\perp l} v^2 +b^2\delta\eta_{lll}-\frac{a^2b^2\delta\lambda(\chi_\perp+\lambda_\perp+\delta\lambda+\delta\chi) }{4\varepsilon H^\parallel_\parallel(V,\xi)}\;,\\
	&&U_1^\mu\xi_\mu =b^3\delta\eta_{lll\perp} +a^2b\delta\lambda+(\delta\eta_{\perp l}+\delta\eta_{lll})b v^2-\frac{a^2b\delta\lambda[(\chi_\perp+\lambda_\perp)v^2 +b^2(\delta\lambda+\delta\chi)]}{4\varepsilon H^\parallel_\parallel(V,\xi)}\;,\\
	&&U_2^\mu l_\mu =b\left (\frac{\delta\chi}{3}+\delta \eta_{lll}\right )+\frac{(\chi_\perp-\eta_{ll})b}{3}-\frac{a^2b (\chi+\lambda_\perp) (\chi_\perp+\lambda_\perp+\delta\lambda+\delta\chi) }{4\varepsilon H^\parallel_\parallel(V,\xi)}\;,\\
	&&U_2^\mu\xi_\mu =b^2\left (\frac{\delta\chi}{3}+\delta \eta_{lll}\right )+\frac{(\chi_\perp-\eta_{ll})v^2}{3}-\frac{a^2 (\chi+\lambda_\perp) [(\chi_\perp+\lambda_\perp)v^2 +b^2(\delta\lambda+\delta\chi)] }{4\varepsilon H^\parallel_\parallel(V,\xi)}\;,\\
	&&H^\parallel_\parallel(V,\xi)=\frac{3\chi a^2+\lambda_\perp v^2+\delta\lambda\, b^2}{4\varepsilon}\;.
	\eea
	\eml
	Thus, the matter sector contains 10 overall roots that must obey (C1) and (C2), with 4 coming from 
	\be
	\label{root1}
	A^2=\left (\lambda_\perp a^2-\eta_\perp v^2+\delta\eta_{\perp l} b^2\right )^2=0\;,
	\ee
	and the remaining 6 from \footnote{Note that in the product $H^\parallel_\parallel(V,\xi)\left (U_1^\mu l_{\mu}U_2^\nu \xi_{\nu}-U_1^\mu \xi_{\mu}U_2^\nu l_{\nu}\right )$, the term with denominator $H^\parallel_\parallel(V,\xi)$ cancels as expected.}
	\bea
	\label{root2}
	H^\parallel_\parallel(V,\xi)\left [A^2+A\left (U_1^\mu l_{\mu}+U_2^\mu \xi_{\mu}\right )+U_1^\mu l_{\mu}U_2^\nu \xi_{\nu}-U_1^\mu \xi_{\mu}U_2^\nu l_{\nu}\right ]=0\;.
	\eea
	
	\section{\MakeUppercase{Causility conditions in  the general and specific cases}}\label{app:specific-causaility}
	
	The polynomial in \eqref{root2} is of power 3 in $a^2$. Since for causality we must obtain the roots in the form of $\varrho=\frac{a^2}{v^2}$, then we may rewrite \eqref{root2} as
	\be
	\label{general_root}
	4\varepsilon H^\parallel_\parallel(V,\xi)\left [A^2+A\left (U_1^\mu l_{\mu}+U_2^\mu \xi_{\mu}\right )+U_1^\mu l_{\mu}U_2^\nu \xi_{\nu}-U_1^\mu \xi_{\mu}U_2^\nu l_{\nu}\right ]=p(\varrho)v^6\,.
	\ee  
	We have multiplied \eqref{root2} by $4\varepsilon$ to eliminate it from the denominator in $H^\parallel_\parallel(V,\xi)$. Note that $p(\varrho)$ is a cubic polynomial in $\varrho$.
	Since $l^\mu=\Delta^{\mu\nu}l_\nu$ and $v^\mu=\Delta^{\mu\nu}\xi_\nu$ are vectors orthogonal to $u^\mu$, $\Delta^{\mu\nu}$ define an inner product between them and, thus, we can apply the Cauchy-Schwarz inequality to write $b=l_\mu v^\mu=\kappa \,v$, where $\kappa\in[-1,1]$ depending on the root $\xi$ and the vector $l$.  Causality of the 6 roots of \eqref{general_root} follows from the statement:
	
	\begin{statement}[The general case]\label{st:general-case}
		Let $p(\varrho)$ be defined by means of Eq.\\ \eqref{general_root} and let us write it as
		\be
		\label{polynomial_p}
		p(\varrho)=\alpha_3\varrho^3+\alpha_2\varrho^2+\alpha_1\varrho+\alpha_0.
		\ee
		Assume that 
		\begin{equation}\label{A2}
			\alpha_3>0\;.
		\end{equation}
		Then, causality requires that
		\bml
		\label{general_cond}
		\bea
		&&p(\varrho)>0,\quad \forall \varrho\ge 1\label{general_cond_1}\\
		&&p(\varrho)<0,\quad \forall \varrho<0\label{general_cond_2},\\
		&&18\alpha_0\alpha_1\alpha_2\alpha_3-4\alpha_2^3\alpha_0+\alpha_2^2\alpha_1^2-4\alpha_3\alpha_1^3-27\alpha_3^2\alpha_0^2\ge0,\label{general_cond_3}
		\eea
		for all $\kappa\in[-1,1]$. 
		\eml
	\end{statement}
	\begin{proof}
		First, we must ensure that all real roots of $p(\varrho)$ lie on the range $[0,1)$ as demanded by \eqref{C2}. From \eqref{A2}, we must ensure that $p(\varrho)$ is positive for all $\varrho\ge 1$ [condition \eqref{general_cond_1}] and negative for all $\varrho<0$ [condition \eqref{general_cond_2}. This guarantees that the real roots can only occur in this desired range, and thus \eqref{C2} is satisfied. As for \eqref{C1}, the roots are real if the discriminant of the cubic polynomial \eqref{polynomial_p} is greater than or equal to zero, which leads to the condition \eqref{general_cond_3}. 
			\end{proof}
			In what follows, we present causality conditions for two more specific cases. The first case is when the anisotropy appears only in $\mcE^{(1)}$, $\mcP^{(1)}_l$, and $\mcP^{(1)}$:
			\begin{statement}[Anisotropy in the $\chi$'s]\label{st:chi-causality}
				Consider the anisotropic conformal fluid theory defined by the energy-momentum tensor in \eqref{1} and supplemented with Eqs.\ \eqref{Definitions} with the choices $\lambda_\perp=\lambda_l=\lambda,$ and $\eta_\perp=\eta_l=\eta_{ll}=\eta$. Then, the corresponding EOM are causal under assumption \eqref{A1} if, and only if, condition \eqref{condition_lambda} applies together with 
				\bml
				\label{tensor_1_conditions}
				\bea
				&&\kappa^2 \delta \chi  \lambda +\chi  (4 \eta +\lambda -\chi )+(\lambda +\chi ) \chi _\perp\ge 0\label{tensor_1_conditions-2}\\
				&&\kappa^4 \delta \chi ^2 \lambda ^2-2 \kappa^2   \lambda  \chi \delta \chi  (-4 \eta +\lambda +\chi )+2 (\lambda +\chi )
				\chi _\perp \left[\kappa^2 \delta \chi  \lambda +\chi  (4 \eta +\lambda -\chi )\right]\nonumber\\
				&&+\chi  \left[16 \eta ^2 \chi +8
					\eta  \left(2 \lambda ^2+\lambda  \chi -\chi ^2\right)+\chi  \left(-3 \lambda ^2-2 \lambda  \chi +\chi
				^2\right)\right]+(\lambda +\chi )^2 \chi _\perp^2\ge0,\label{tensor_1conditions-3}\\
				&&\chi\ge 4\eta-\kappa^2 \delta\chi,\label{tensor_1conditions-4}\\
				&&-\kappa^2  \lambda\delta \chi +\chi  (-4 \eta +5 \lambda +\chi )-(\lambda +\chi ) \chi _\perp>0,\label{tensor_1conditions-5}\\
				&&-2 \kappa^2 \delta \chi  \lambda -4 \eta  \lambda -12 \eta  \chi +7 \lambda  \chi -3 (\lambda +\chi ) \chi _\perp+3 \chi^2>0,\label{tensor_1conditions-6}
				\eea
				\eml
				for all $\kappa^2\in[0,1]$.
			\end{statement}
			\begin{proof} 
				The determinant $M$ in \eqref{M} can be rewritten as
				\bea
				M&=&\frac{3\chi\lambda^4}{4\varepsilon}\prod_{a=1,\pm}(a^2-\tau_a v^2)^{n_a},
				\eea
				where $n_1=3$, $n\pm=1$, and
				\bml
				\bea
				\tau_1&=&\frac{\eta}{\lambda},\label{tau_2}\\
				\tau_\pm&=&\frac{\alpha \pm\sqrt{\beta} }{6 \lambda \chi }\label{tau_3},\\
				\alpha&=&\kappa^2 \delta \chi  \lambda +\chi  (4 \eta +\lambda -\chi )+(\lambda +\chi ) \chi _\perp,\\
				\beta&=&\kappa^4 \delta \chi ^2 \lambda ^2-2 \kappa^2   \lambda  \chi \delta \chi  (-4 \eta +\lambda +\chi )+2 (\lambda +\chi )
				\chi _\perp \left[\kappa^2 \delta \chi  \lambda +\chi  (4 \eta +\lambda -\chi )\right]\nonumber\\
				&&+\chi  \left[16 \eta ^2 \chi +8
					\eta  \left(2 \lambda ^2+\lambda  \chi -\chi ^2\right)+\chi  \left(-3 \lambda ^2-2 \lambda  \chi +\chi
				^2\right)\right]+(\lambda +\chi )^2 \chi _\perp^2.   
				\eea
				\eml
				The matter sector has 2 roots for \eqref{tau_2} with multiplicity 3 each and 4 roots for \eqref{tau_3} (two for each $\tau_\pm$), a total of 10 roots.  
        Now, Eqs.\ \eqref{C1} and \eqref{C2} are observed if, and only if, $0\le \tau_a<1$. For $\tau_1$ it is guaranteed by \eqref{A1} together with \eqref{condition_lambda}. As for $\tau_\pm$, it needs to be real, i.e., $\beta\ge0$, what corresponds to \eqref{tensor_1_conditions-2}, $\tau_-\ge 0$, and $\tau_+<1$. For $\tau_-\ge 0$, we need that $\alpha\ge0$ [condition \eqref{tensor_1conditions-3}] together with $\alpha^2-\beta\ge0$ [condition \eqref{tensor_1conditions-4}]. As for $\tau_+<1$, it corresponds to $6 \lambda \chi-\alpha\ge 0$ [condition \eqref{tensor_1conditions-5}] together with $(6 \lambda \chi-\alpha)^2-\beta>0$, i.e., condition \eqref{tensor_1conditions-6}. All the above conditions must be valid for all values of $\kappa\in[-1,1]$, i.e., for all possible values of the product $b=l_\mu v^\mu=\kappa v$.
			\end{proof}
			Finally, a much simpler case is given below:
			\begin{statement}[The shearless case]\label{st:shearless}
				Consider the shearless anisotropic fluid ($\eta_\perp=\eta_l=\eta_{ll}=0$) with $\delta\chi=0$ described by the energy-momentum tensor \eqref{EMT} and supplemented by \eqref{Definitions}. The theory is causal if \eqref{A1} is satisfied.
			\end{statement}
			\begin{proof}
				In this particular case, the determinant in \eqref{M} becomes
				\bea
				M&=&\frac{3\chi \lambda_\perp^4 a^6}{4\varepsilon}\left( a^2-\frac{1}{3}v^2\right)^2=0. 
				\eea
				Note that there is 1 causal root $a=\xi_\mu u^\mu=0$ because in this case $\xi_\mu \xi^\mu=-a^2+v^2=v^2>0$, with multiplicity 6, and 2 roots $a^2= v^2/3$, also causal since $0<\tau=1/3<1$, with multiplicity 2 each, completing the total 10 roots from the matter sector. Assumption (A1) guarantees that the determinant $M$ is not trivially zero, which would give any $\xi_\mu$ as a possible solution.
			\end{proof}
			
			\section{\MakeUppercase{A causal example}}\label{app:causal-example}
			In this appendix, we use statement \ref{st:general-case} to show that the set of parameters \eqref{example}, which are reproduced below for convenience, are causal
			\[
				\eta_\perp=\eta\,,\quad\eta_l=\frac{2\eta}{3}\,,\quad\eta_{ll}=\frac{5\eta}{6}\,,\quad\lambda_\perp=\frac{13\eta}{2}\,,\quad\lambda_l=6\eta\,,\quad\chi=5\eta\,,\quad\chi_\perp=\frac{11\eta}{2}\,,\quad\chi_l=\frac{16\eta}{3}\;.
			\]
			The above parameters satisfy \eqref{A1} and \eqref{Condition_root_A} by construction, and we show that satisfies the conditions of statement \ref{st:general-case} as well. For simplicity, we take the overall factor $\eta^3/216$ out of the definition of $p(\varrho)$ in \eqref{general_root} to obtain
			\be
			4\varepsilon H^\parallel_\parallel(V,\xi)\left [A^2+A\left (U_1^\mu l_{\mu}+U_2^\mu \xi_{\mu}\right )+U_1^\mu l_{\mu}U_2^\nu \xi_{\nu}-U_1^\mu \xi_{\mu}U_2^\nu l_{\nu}\right ]=\frac{\eta^3}{216}p(\varrho)v^6\;.
			\ee 
			Then, the coefficients in $p(\varrho)$ are given by
			\bml
			\bea
			\alpha_0 &=& -364 - 1623 \kappa^2 + 1674 \kappa^4 - 119 \kappa^6<0,\quad \forall\; \kappa^ 2\in[0,1]\\
			\alpha_1 &=& 16224 + 19601 \kappa^2 - 19925 \kappa^4>0,\quad \forall\;\kappa\in[0,1]\\
			\alpha_2 &=& -18 (7254 - 203 \kappa^2)<0,\quad \forall\;\kappa\in[0,1]\\
			\alpha_3&=&126360\;,
			\eea 
			\eml
			where $\kappa$ is defined through $l_\mu v^\mu = \kappa v$ (see Appendix \ref{app:specific-causaility}).
			For $\varrho < 0$, we have
			\be
			p(\varrho) = -\left(\abs{\alpha_0}+\alpha_1\abs{\varrho}+\abs{\alpha_2}\varrho^2+\alpha_3\abs{\varrho}^3\right)<0\;,
			\ee
			and, thus, \eqref{general_cond_2} is verified.
			On the other hand, for $\varrho \ge 1$,
			\be
			p(\varrho)=\alpha_0+\alpha_1\varrho+(\alpha_2+\alpha_3\varrho^2)\varrho\ge \alpha_0+\alpha_1+\alpha_2+\alpha_3\;,
			\ee
			since $\alpha_1,\alpha_3>0$. However,
			\be
			\alpha_0+\alpha_1+\alpha_2+\alpha_3 = 11648 + 21632 \kappa^2 - 18251 \kappa^4 - 119 \kappa^6>0,\quad \forall\, \kappa^2\in[0,1],
			\ee
			and, therefore, \eqref{general_cond_1} is also verified. Finally, 
			\bea
			&&18\alpha_0\alpha_1\alpha_2\alpha_3-4\alpha_2^3\alpha_0+\alpha_2^2\alpha_1^2-4\alpha_3\alpha_1^3-27\alpha_3^2\alpha_0^2=324 \left (2421746461615104 - 6266597114164608 \kappa^2
			\right .\nonumber\\
			&& 
			+24310721163158820 \kappa^4- 50747526702105948 \kappa^6 + 
			59336411451755437 \kappa^8 - 40187158939087070 \kappa^{10}\nonumber\\
			&& \hspace{3cm}  \left . +  12398536143066361 \kappa^{12}\right )>0\;,\quad\forall\, \kappa^2\in[0,1]\;.
			\eea
			Hence, \eqref{general_cond_3} is also verified and the set of transport parameters \eqref{example} is causal.
			\section{\MakeUppercase{Stability of the causal example \eqref{example}}}\label{app:stability-example}
			
			In this appendix, we examine the stability of the sound channel \eqref{sound} for the causal set of parameters \eqref{example}.
			In this case, up to an overall constant and, because $\bar{\eta}>0$, by performing the changes $\Gamma\to \Gamma/\bar{\eta}$ and $k^i\to k^i/\bar{\eta}$  we obtain that Eq.\\ \eqref{sound} becomes
			\be
			\label{sound_C-a}
			a_0\Gamma^6+a_1\Gamma^5+a_2\Gamma^4+a_3\Gamma^3+a_4\Gamma^2+a_5\Gamma+a_6=0\;,
			\ee
			where
			\bml
			\label{coefficients}
			\bea
			a_0&=&21060\;,\\
			a_1&=&10962\;,\\
			a_2&=&6 \left[315 + f_1(x^ 2) k^2\right]\;,\\
			a_3&=&108 + f_2(x^2) k^2\;,\\
			a_4&=&\frac{k^2}{2} \left[f_3(x^2) + f_4(x^2) k^2\right],\\
			a_5&=&36 k^2 + f_5(x^2) k^4 \;,\\
			a_6&=&k^4 \left[f_6(x^2) + f_7(x^2) k^2\right]\;.
			\eea
			\eml
			We have defined the functions
			\bml
			\label{functions}
			\bea
			f_1(x^2)&=&3498 - 70 x^2\;,\\
			f_2(x^2)&=&7248 - 270 x^2\;,\\
			f_3(x^2)&=&1680 - 36 x^2\;,\\
			f_4(x^2)&=&5548 + 6564 x^2 - 6464 x^4\;,\\
			f_5(x^2)&=&485 + 556 x^2 - 553 x^4\;,\\
			f_6(x^2)&=&24 + 30 x^2 - 30 x^4\;,\\
			f_7(x^2)&=&78 + 306 x^2 - 310 x^4 + 22 x^6\;.
			\eea
			\eml
			One may verify that all functions in \eqref{functions} are positive for all $x^2\in[0,1]$, which makes all $a_I$'s in \eqref{coefficients} ($I=0,\,\cdots\,,6$) to also be positive. Thus, pure real roots are in fact negative. As for imaginary roots, we apply the Routh-Hurwitz criterion (RHC) \cite{gradshteyn2007}, which in this case requires us to compute the following table:
			\be
			\begin{array}{|c|c|c|c|c|}
				\hline
				a_0 & a_2 & a_4 & a_ 6 & 0 \\
				\hline
				a_1 & a_3 & a_5 & 0    & 0 \\
				\hline
				b_1 & b_2 & b_3 & 0    & 0 \\
				\hline
				c_1 & c_2 & 0   & 0    & 0 \\
				\hline
				d_1 & d_2 & 0   & 0    & 0 \\
				\hline
				e_1 & 0   & 0   & 0    & 0 \\
				\hline
			\end{array}
			\ee
			where
			$b_i=(a_1a_{2i}-a_0a_{2i+1})/a_1$, $c_i=(b_1a_{2i+1}-a_1 b_{i+1})/b_1$, $d_i=(c_1 b_{i+1}-b_1c_{i+1})/c_1$, and $e_1=(d_1c_2-c_1 d_2)/d_1$. Since $a_I>0$ for $I=0\,,\cdots\,,6$, then $\Re(\Gamma)<0$ if, and only if, $b_1>0$, $c_1>0$, $d_1>0$, and $e_1>0$. Since $a_1>0$, then it is enough to obtain
			\bml
			\label{RHconditions}
			\bea
			a_1b_1&=&324 \left[56925 + (238974 + 3340 x^2) k^2\right]>0\;,\\
			b_1 c_1&=&\frac{36}{203} \left[1024650 + 
			18 g_1(x^2) k^2 + g_2(x^2) k^4\right]\;,\\
			b_1c_1d_1&=&\frac{36k^ 2}{203}\left[6147900 g_3(x^2)+27 g_4(x^ 2)k^2+6 g_5(x^ 2)k^4+g_6(x^2)k^6\right]\;,\\
			b_1c_1d_1e_1&=&\frac{36k^4}{203} \left[221324400 g_7(x^2)+972 g_8(x^2)k^2+27g_9 (x^2)k^4+6g_{10}(x^2)k^6
			+g_{11}(x^2)k^8\right],   
			\eea
			\eml 
			where
			\bml
			\bea
			g_1(x^2)&=&1412154 - 77159 x^2\;,\\
			g_2(x^2)&=&174806118 - 143563237 x^2 + 133959417 x^4\;,\\
			g_3(x^2)&=&35 - 3 x^2\;,\\
			g_4(x^2)&=&422612235 - 188238658 x^2 + 129974883 x^4\;,\\
			g_5(x^2)&=&20839149333 - 15122827103 x^2 + 12589851858 x^4 + 623807058 x^6\;,\\
			g_6(x^2)&=&219646045656 + 96361454524 x^2 - 432604607986 x^4 +  620066970164 x^6\nonumber\\
			&& - 290471866054 x^8\;,\\
			g_7(x^2)&=&23 - 18 x^2 + 15 x^4\;,\\
			g_8(x^2)&=&362387218 - 297049991 x^2 + 243855688 x^4 - 5179815 x^6\,\\
			g_9(x^2)&=&216917135055 - 156807618144 x^2+ 105968706122 x^4+ 35462079400 x^6\nonumber\\
			&& - 13719895953 x^8\;,\\
			g_{10}(x^2)&=&5000654914056 - 3108012289024 x^2 + 1415965400581 x^4 + 1798665419999 x^6 \nonumber\\
			&& - 134955905301 x^8- 625170319671 x^{10}\;,\\
			g_{11}(x^2)&=&20097012270600 - 82381572326488 x^2 + 329458308450878 x^4 - 
			682695134979400 x^6 \nonumber\\
			&&+ 795041858933812 x^8 - 532773739193376 x^{10} + 161024463542854 x^{12}\;.
			\eea
			\eml
			Since all 11 functions $g$ are greater than zero for all $x^ 2\in[0,1]$, then all equations in \eqref{RHconditions} are positive, the RHC are verified and the system is stable in the LRF. Since we proved in Appendix\ \ref{app:causal-example} that the system under these parameters is causal, then the result in \cite{Gavassino:2021owo} ensures linear stability in any boosted frame.
			
			Just for the sake of illustration, let us verify stability in the boosted homogeneous frame by taking the changes \eqref{B1} in \eqref{sound_C-a} and then taking $k^i=0$. This will lead us to the roots $\Gamma=0$ of multiplicity 3 and the roots of X,
			\be
			\beta_0 (\gamma\Gamma)^3+\beta_1(\gamma\Gamma)^2+\beta_2\gamma\Gamma+\beta_3=0\;,
			\ee 
			where
			\bml
			\bea
			\beta_0&=&2 (1384 + 1698 x^2 - 1461 x^4 - 11 x^6)>0,\quad\forall\;x^ 2\in[0,1]\;,\\
			\beta_1&=&4199 + 826 x^2 - 553 x^4>0,\quad\forall\;x^ 2\in[0,1]\;,\\
			\beta_2&=&6 (179 + 8 x^2 - 5 x^4)>0,\quad\forall\;x^ 2\in[0,1]\;,\\
			\beta_3&=&72\;.
			\eea
			\eml
			Since all coefficients $\beta_{0,1,2,3}$ are positive, then there is only 1 negative pure real root for the polynomial, as desired. As for the complex roots, the remaining RHC to be computed is
			\be
			\beta_1\beta_2-\beta_0\beta_3=6 (718405 + 140694 x^2 - 78310 x^4 - 8290 x^6 + 2765 x^8)>0\;,
			\ee
			which is greater than zero for all $x^ 2\in[0,1]$. Thus, linear stability is also verified in the homogeneous boosted frame.
			\section{\MakeUppercase{The unphysical choice of $l$ in the Bjorken flow}}\label{app:wrong}
			
			Here, we repeat the study of Bjorken flow from Sec.\ \ref{sec:bjorken}, with an alternative choice of spacelike anisotropy vector, 
			$
			l = \pdv{x}
			$.
			The fluxes of \eqref{Definitions} in this case are
			\begin{subequations}
				\bea
				\mathcal{E}^{(1)}&=&\tilde{\chi}^3T^3\left(\inv{\tau}+\frac{3\dot{T}}{T}\right)
				\\
				\mathcal{P}_l^{(1)}&=&T^3\left(\frac{2\tilde{\eta}_{ll}+\tilde{\chi}_l}{3\tau}+\frac{3\tilde{\chi}_l\dot{T}}{T}\right)
				\\
				\mathcal{P}_\perp^{(1)}&=&T^3\left(\frac{\tilde{\chi}_\perp-\tilde{\eta}_{ll}}{3\tau}+\frac{3\tilde{\chi}_\perp\dot{T}}{T}\right)
				\\
				\pi_\perp^{\mu\nu}&=&-2\eta_\perp T^3 \mathrm{diag}\left(0,0,1/\tau,-1/\tau^3\right),\\
				W_{\perp l}^\mu&=&0,\\
				W_{\perp u}^\mu&=&0,\\
				M&=&0,
				\eea
			\end{subequations}
			which gives rise to the following EOM
			\begin{eqnarray}
				&&9\tilde{\chi}\frac{\tau^2\ddot{T}}{T} + 18\tilde{\chi}\frac{\tau^2\dot{T}^2}{T^2}+ \left(\frac{3\tau(6\tilde{\chi}+\tilde{\chi}_\perp)}{T}+12\tau^2\right)\dot{T}+4\tau T + \tilde{\chi}_\perp - 3\tilde{\eta}_\perp -\tilde{\eta}_{ll} = 0\,.
			\end{eqnarray}
			The above can be expressed in terms of $w=T\tau$ and $f(w)=\frac{\tau}{w}\dv{w}{\tau}$ as
			\begin{eqnarray}
				\frac{9\tilde{\chi}}{4}f(w)^2+
				w f(w)\left(1+\frac{3}{4}\tilde{\chi}f'(w)\right)
				- \frac{\tilde{\chi}-15\tilde{\chi}_\perp}{4}f(w) +\frac{18\tilde{\chi}-2\tilde{\chi}_\perp-3\tilde{\eta}_\perp-\tilde{\eta}_{ll}}{12}-\frac{2w}{3} = 0\,.
			\end{eqnarray}
			The pressure anisotropy reads
			\begin{equation}
				\mcA = \frac{2}{3}\frac{\tilde{\chi}_\perp-\tilde{\chi}_l}{w} \left(f-\frac{2}{3}\right) - \frac{3\tilde{\eta}_{ll}}{w}\,.
			\end{equation}
			As in Sec. \ref{sec:bjorken}, $\mcA$ has an ``off-shell'' contribution, which cannot be found in the isotropic conformal BDNK.  However, the first-order term is negative, in contrast to the isotropic case. 
			The late-time expansion for $T$ is
			\begin{equation}
				T = \frac{\Lambda}{(\Lambda\tau)^{1/3}}\left(1-\frac{\tilde{\eta}_{ll}+3\tilde{\eta}_{\perp }}{8(\Lambda\tau)^{2/3}}
				-\frac{(\tilde{\eta}_{ll}+3\tilde{\eta}_{\perp})(5\tilde{\chi}-\tilde{\chi}_\perp)}{64(\Lambda\tau)^{2/3}}+\cdots\right)\,,
			\end{equation}
			and for $f$  is
			\begin{equation}
				f(w) = \frac{2}{3} + \frac{\tilde{\eta}_{ll}+3\tilde{\eta}_{\perp}}{12w}+\frac{(\tilde{\eta}_{ll}+3\tilde{\eta}_{\perp})(5\tilde{\chi}-\tilde{\chi}_\perp)}{48w^2} + \order{\inv{w^3}}\,.
			\end{equation}
			We may consider the following linear perturbation at late times
			\begin{equation}
				f(w) = \frac{2}{3} + \frac{\tilde{\eta}_{ll}+3\tilde{\eta}_{\perp}}{12w} + \delta f(w)\,,
			\end{equation}
			which up to the first order in perturbation in late times is
			\begin{equation}
				\delta f(w) \sim \exp(-\frac{2w}{\tilde{\chi}}) w^{\frac{\tilde{\eta}_{ll}+3\tilde{\eta}_{\perp}+2(\tilde{\chi}_\perp+\tilde{\chi}_l)}{4\tilde{\chi}}}\,.
			\end{equation}
			The numerical attractor can be found from the following initial condition
			\begin{equation}
				f(w\ll 1) = \frac{7}{9} +\frac{\tilde{\chi}_l-\tilde{\chi}_\perp}{36\tilde{\chi}}+ \frac{\sqrt{12\left(\tilde{\eta}_{ll}+3\tilde{\eta}_\perp\right)\tilde{\chi}+\left(\tilde{\chi}_\perp+\tilde{\chi}_l\right)^2}}{18\tilde{\chi}}\,,
			\end{equation}
			and the slow-roll attractor is
			\begin{equation}
				f(w)_{\rm slow roll} = \frac{7}{9} +\frac{\tilde{\chi}_l-\tilde{\chi}_\perp}{36\tilde{\chi}} -\frac{2w}{9\tilde{\chi}}+\frac{\sqrt{\left(4w-(\tilde{\chi}_l+\tilde{\chi}_\perp)\right)^2+12\left(3\tilde{\eta}_{\perp}+\tilde{\eta}_{ll}\right)\tilde{\chi}}}{18\tilde{\chi}}
			\end{equation}
			Neglecting the power counting argument, we find $\nabla\cdot S_{\rm off}$ is initially negative if
			\begin{equation}
				\chi_\perp > \eta_{ll} + 3\eta_{\perp} > 0\,.
			\end{equation}
			In the isotropic limit, the above reproduces the stability condition
			\[
				\chi > 4\eta\,.
			\]
			The condition $\nabla\cdot S < 0$ at $w=0$ is equivalent to 
			\begin{equation}
				\frac{2}{3} < f(0) < 1\,,
			\end{equation}
			which prevents early reheating for the attractor solution. 
   \bibliographystyle{apsrev4-1}
			\bibliography{References}
\end{document}